\documentclass[iop]{emulateapj}%

\usepackage{natbib}
\usepackage{amsmath}
\usepackage{multirow}

\newcommand{\degree}{\ensuremath{^\circ}}
\newcommand{\V}[1]{\boldsymbol{#1}}

\begin{document}

\title{CHARA/MIRC observations of two~M~supergiants in Perseus OB1:\\
Temperature, Bayesian modeling, and compressed sensing imaging}

\author{F. Baron\altaffilmark{1,9}, J. D. Monnier\altaffilmark{1}, L.L. 
Kiss\altaffilmark{2,3}, H.~R.~Neilson\altaffilmark{4}, M.~Zhao\altaffilmark{5}, 
M.~Anderson\altaffilmark{1}, A. Aarnio\altaffilmark{1}, 
E.~Pedretti\altaffilmark{6}, N.~Thureau\altaffilmark{6}, T.A. ten
 Brummelaar\altaffilmark{7}, S.T. Ridgway\altaffilmark{8}, H.A. 
McAlister\altaffilmark{7,9}, J.~Sturmann\altaffilmark{7}, 
L.~Sturmann\altaffilmark{7}, N.~Turner\altaffilmark{7}}
\affil{\altaffilmark{1}Department of Astronomy, University of Michigan, 918 
Dennison Building, Ann Arbor, MI48109-1090, USA;}
\email{baron@phy-astr.gsu.edu}
\affil{ \altaffilmark{2} Sydney Institute for Astrophysics, School of Physics, 
University of Sydney, NSW 2006, Australia;}
\affil{ \altaffilmark{3} Konkoly Observatory, Hungarian Academy of Sciences, 
Budapest, Hungary;}
\affil{ \altaffilmark{4} Department of Physics \& Astronomy, East Tennessee 
State University, Box 70652, Johnson City, TN 37614, USA;}
\affil{ \altaffilmark{5} Department of Astronomy \& Astrophysics, Penn State 
University, University Park, PA 16802 , USA;}
\affil{\altaffilmark{6}  Department of Physics \& Astronomy, University of St. 
Andrews, Scotland, UK;}
\affil{ \altaffilmark{7} The CHARA Array, Georgia State University,  P.O. Box 
3965, Atlanta, GA 30302-3965, USA;}
 \affil{\altaffilmark{8}National Optical Astronomy Observatory, Tucson, AZ 
85726-6732 USA.}
\affil{\altaffilmark{9}  CHARA and Department of Physics \& Astronomy, Georgia 
State University, P. O. Box 4106, Atlanta, GA 30302-4106, USA.}

\begin{abstract}
Two red supergiants of the Per OB1 association, RS~Per and T~Per, have been
observed in H~band using the MIRC instrument at the CHARA array. The data 
show clear evidence of departure from circular symmetry. We present here new 
techniques specially developed to analyze such cases, 
based on state-of-the-art statistical frameworks. 

The stellar surfaces are first modeled as limb-darkened discs based on SATLAS 
models that fit both MIRC interferometric data and publicly available 
spectrophotometric data. Bayesian 
model selection is then used to determine the most probable number of spots. 
The effective surface temperatures are also determined and give further 
support to the recently derived hotter temperature scales of red supergiants.
The stellar surfaces are reconstructed by our model-independent imaging code 
SQUEEZE, making use of its novel regularizer based on Compressed Sensing
theory. We find excellent agreement between the model-selection results and the 
reconstructions. Our results provide evidence for the presence of near-infrared 
spots representing about 3-5\% of the stellar flux. 
\end{abstract}


\section{Introduction}

Red supergiants (RSGs) represent an important but still poorly characterized
evolutionary phase of massive stars. As He-burning evolved stars, their
surfaces present very cool effective temperatures between 3400 and 4100 K
(spectral type of late-K to M) and average luminosities 20~000 to 300~000
$L_{\odot}$ \citep{Levesque2005, Levesque2006}. RSGs are amongst the largest
stars, with radii up to 1500 $R_\odot$ and masses in the 10-25 $M_{\odot}$
range \citep{Levesque2005, Massey2008, Levesque2010}. The resulting low
gravity has the consequence that material from the outer envelope can easily
escape to the interstellar medium, giving rise to circumstellar envelopes
\citep{Danchi1994} and very significant mass-loss rates ranging between
$10^{-8}$ to $10^{-4} M_{\odot}$ $\text{yr}^{-1}$ \citep{Massey2005,
  Verhoelst2009, Sargent2011}. The actual mechanism of the mass loss is still 
unknown but is likely to involve the combined effects of turbulent pressure 
from large convection 
cells and radiation pressure on molecular lines \citep{Josselin2007}, as well 
as 
stellar magnetism \citep{Grunhut2010, Auriere2010} and Alfv{\'e}n winds
\citep{Cuntz1997, Airapetian2010}. Through these processes, RSGs are key
agents of nucleosynthesis and chemical enrichment of the Galaxy.

In the last decade, photometric and spectroscopic observations have allowed
several breakthroughs in our understanding of their dynamical convective
patterns. RSGs have long been known for their semi-regular short term optical
variations with periods of the order of hundreds of days \citep{Kiss2006}. 
Historically, this
variability was usually attributed to radial pulsation, and while this path is
still being investigated \citep{Yang2012}, the full explanation is now thought
to involve variation of a few large granules on the surface of the RSGs. Such
huge convection cells are suggested by theory and simulations
\citep{Schwarzschild1975, Stothers2010}, and supported by recent spectroscopic
observations that detected large amounts of material moving through the
photosphere of RSGs \citep{Josselin2007, Gray2008, Ohnaka2009, Ohnaka2011, 
Ohnaka2013}. 

%
Modeling these convection effects requires a precise determination of the
temperature of the RSGs. The last decade of measurements of effective
temperatures of RSGs seemed to indicate much cooler temperatures than
predicted by stellar evolutionary theory, until \citet{Levesque2005} used
MARCS stellar atmosphere models with state-of-the art (at the time) treatment
of molecular opacities \citep{Gustafsson1975, Plez2003} to fit
moderate-resolution optical spectrophotometry of Galactic RSGs. They derived a
warmer effective temperature scale for RSGs of Galactic metallicity than
previous studies, in rough agreement with the Geneva evolutionary tracks
\citep{Meynet2003, Levesque2006}. Long-baseline interferometry brings unique 
insights to the study of RSGs. Because optically resolving RSGs by 
interferometry gives direct access to their angular diameters, it is 
complementary to spectrophotometry. In a recently published survey of~$74$ RSGs 
with the Palomar Testbed Interferometer (PTI), \citet{vanBelle2009}
thus derived a slightly warmer temperature scale than that of
\citet{Levesque2005}. 

Convection processes are involved in the creation of hotspots, and determining 
how the formation and evolution of hotspots is correlated to the fundamental
stellar parameters is a difficult task. The presence of spots and large 
convection cells affects the estimation of these fundamental parameters, 
as it leads to short-term photometric variability \citep{Chiavassa2011} in 
addition to long-term effects \citep{Kiss2006}, while also producing photocenter shifts 
that throw off diameter determination \citep{Chiavassa2009}.

High angular resolutions techniques are becoming essential tools to 
understand these hotspots. Using aperture masking at the William Herschel 
Telescope, \citet{Buscher1990} and \citet{Tuthill1997} found bright asymmetries at visible 
wavelengths on the surface of M-supergiants ($\alpha$ Ori $\alpha$~Sco, 
$\alpha$~Her), with timescale variations of order a few months possibly 
explained by the presence
of hotspots. Due to its large angular size, $\alpha$ Ori then became the
best-studied individual RSG in terms of multi-wavelength surface
imaging. Using the COAST interferometer, \citet{Young2000} found a strong
variation in the apparent asymmetry as a function of wavelength, with the
detection of hotspots in the visible, but only featureless disk in
J-band. This has led to the suggestion that the bright spots are unobscured 
regions of elevated temperature, seen
through a geometrically-extended and line-blanketed atmosphere, in which the
features are seen along lines of sight for which the atmospheric opacity has
been reduced as the result of activity (e.g. convection) at the stellar
surface. In H or J band, the continuum opacity is close to minimum in these 
cool atmospheres \citep{Woodruff2009}, one would expect to see the photosphere, 
with 
negligible or no evidence of hotspots at this band. 

However, interferometric observations of AGB stars (somewhat less massive and 
less luminous than RSGs) have revealed that a significant fraction of these 
present strong closure phase signals \citep{Ragland2006}. These signals may be 
explained by unresolved bright spots, though circumstellar emission could not 
ruled out. While the envelopes and dust shells of several RSGs have been 
successfully imaged and shown to be very often asymmetric \citep{Monnier2004a, 
Kervella2011}, resolving actual surface features has proved more difficult. It 
is only recently that \citet{Haubois2009} reported the unambiguous detection of 
two hotspots on $\alpha$ Ori by the IOTA interferometer in H band, while 
\citet{Chiavassa2010a} found a similar number of spots in the same band on 
VX~Sgr using with VLTI/AMBER. These spots are thought to be the imprint of 
giant convection cells based on 3D stellar convection models 
\citep{Freytag2002, Chiavassa2009, Stothers2010, Chiavassa2010b}.

We present in section \ref{sec:obs} of this paper our observations of two RSGs
from the Per OB1 association, T~Per and RS~Per, using the world-leading
resolution of the CHARA Array to resolve their surface in H-band. Then in
Section~\ref{sec:modeling} we attempt to model the stellar surface asymmetries
as spots, and we devise a procedure to determine the probabilities of these
models based on state-of-the-art Bayesian techniques. Using our best estimates
of the stellar diameters, we then derive the linear sizes, bolometric fluxes,
and temperatures of both stars. Finally in
Section~\ref{sec:image_reconstruction} we present model-independent images of
both RSGs obtained with the latest version of the software SQUEEZE and a new
regularizer developed for spotted star reconstruction.

\section{Observations} \label{sec:obs}

\begin{figure}
\centering
\includegraphics[width=0.5\linewidth]{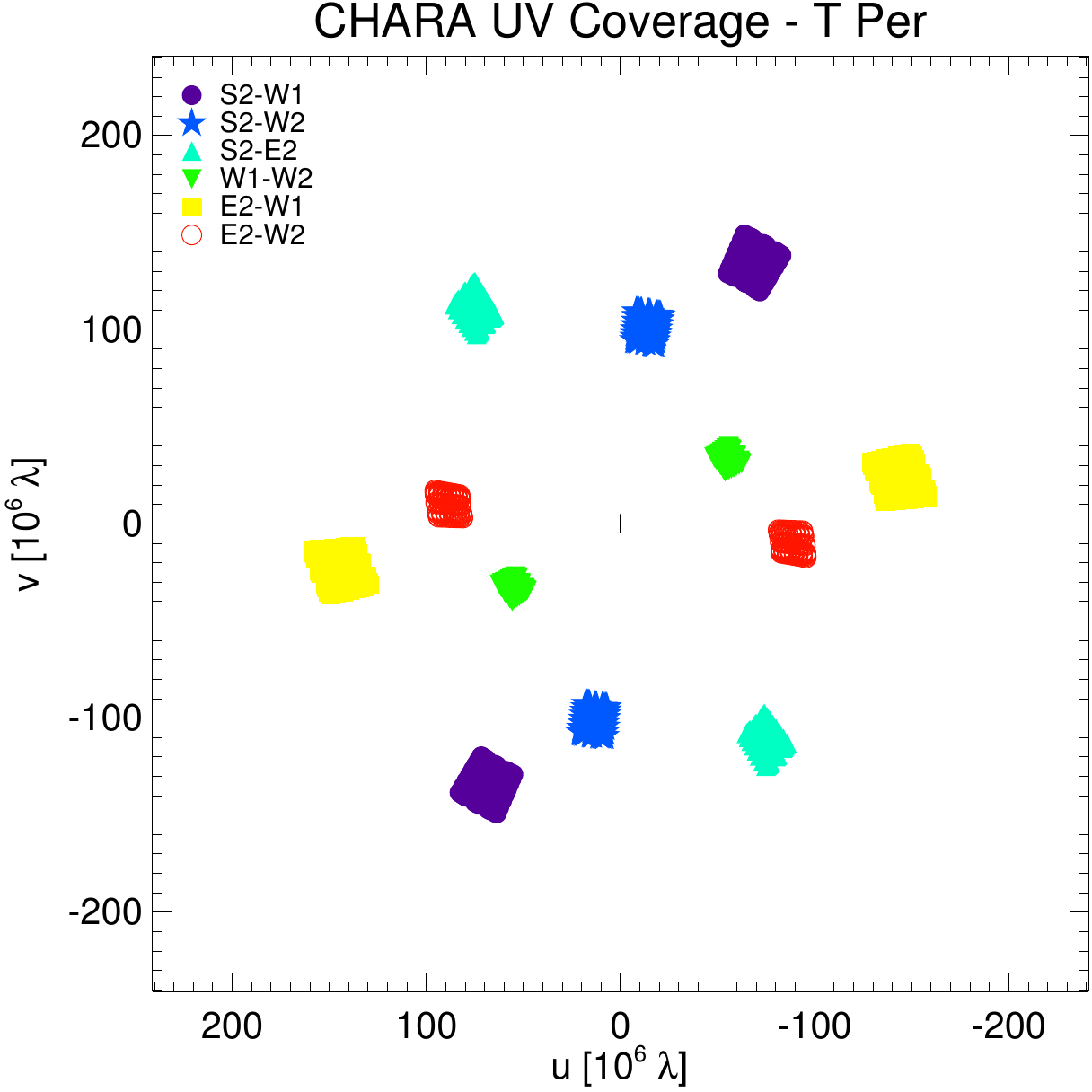}%
\includegraphics[width=0.5\linewidth]{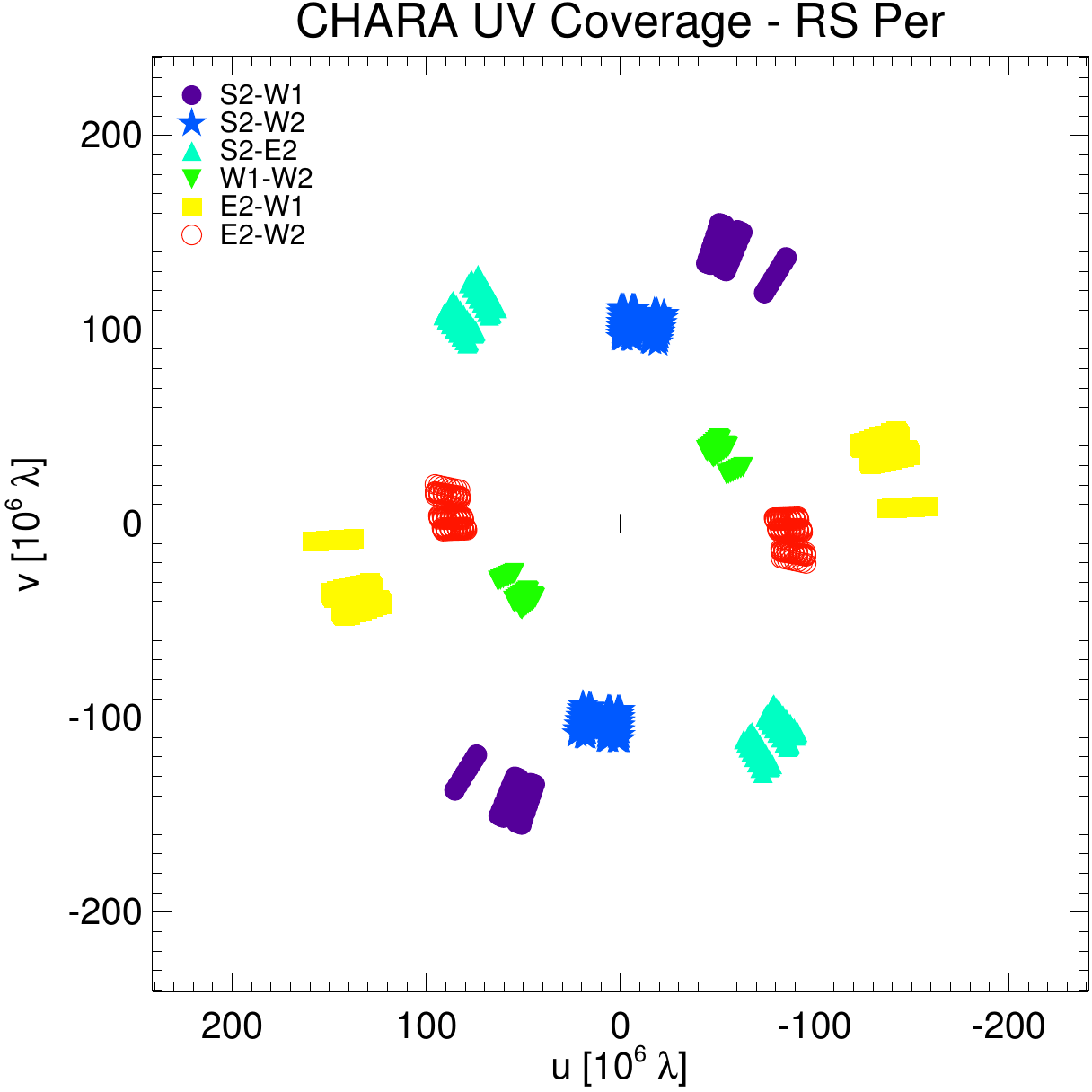}
\caption{Combined ({\it u, v}) coverage of our observations of
T~Per (left) and RS~Per (right). The telescope configuration used
was S2-E2-W1-W2. The radial dispersion is due to the use
of MIRC low spectral mode with 8 channels.}
\end{figure}

\begin{deluxetable}{ccccc}
\centering
\tabletypesize{\scriptsize}
\tablecaption{\label{log} Reduction log for T~Per and RS~Per. $N_{\rm block}$ 
refers to the number of 
data blocks. All calibrated OI-FITS data are available upon request.}
\tablewidth{0pt}
\tablehead{\colhead{Date (UT)} & \colhead{Target} & \colhead{$N_{\rm block}$} & 
\colhead{Calibrators} & \colhead{Flux calibration}}
\startdata
2007 Jul 28 & T~Per & 2 & HD~9022 & Chopper \\
2007 Jul 29 & T~Per & 1 & $\upsilon$~And & Fiber \\
2007 Jul 30 & RS~Per & 1 & $\upsilon$~And & Chopper\\ 
2007 Jul 31 & RS~Per & 1 & 37~And & DAQ \\
2007 Aug 2 & RS~Per & 2 & $\sigma$~Cyg, $\upsilon$~And & DAQ\\
\enddata
\tablecomments{ Calibrator diameters (mas): \\
HD~9022 = $1.05\pm0.02$ \citep{Merand2005}.\\ 
$\upsilon$~And = $1.097\pm0.009$ \citep{Zhao2011}, UD model. \\
37~And = $0.676\pm0.034$, \citet{Kervella2008}. \\
$\sigma$~Cyg = $0.542\pm0.021$ M\'erand 2008, private comm.} 
\label{tab:calibrators}
\end{deluxetable}

\subsection{RS~Per and T~Per}

In the following we will present our observations of two Red Supergiants from 
the Per OB1 association, RS~Per (HD
14488) and T~Per (HD 14142). RS~Per is a firmly established member of the 
$\chi$~Per/NGC~884 cluster, while T~Per lies about 2 degrees North above the 
Double Cluster.
Both are M~supergiants, with RS~Per classed as M4I, and T~Per as M2I.  Based on 
previous results \citep{Gonzalez2000, Slesnick2002, Levesque2005}, T~Per has a
temperature typical of most M~supergiants in the Per OB1 association (average 
temperature in the litterature $T \simeq 3850$ K), while RS~Per is thought to 
be slightly cooler ($T \simeq 3500$~K). It 
is  interesting to note that they both display comparable long photometric 
periods, 
$2500 \pm 460$~days for T~Per, $4200 \pm 1500$ days for RS~Per 
\citep{Kiss2006}, thought to be related to a global pulsation mode. RS~Per also 
displays a shorter period of 
$245$ days.

\subsection{CHARA/MIRC observations}

Our observations were carried out on five nights in July-August 2007 at the 
Georgia State University Center for High Angular Resolution Astronomy (CHARA) 
interferometer
array using the Michigan Infra-Red Combiner (MIRC). The CHARA array, located on
Mount Wilson, consists of six 1~m telescopes. Thanks to its 15~baselines ranging
from 34 m to 331 m, it achieves the highest angular resolution of optical 
interferometers, up to $\sim$0.5 mas in $H$~band \citep{tenBrummelaar2005}.

The Michigan InfraRed Combiner (MIRC) instrument was used to combine the H-band 
light
($1.5$-$1.8 \mu$m) of four CHARA telescopes at low spectral resolution 
(R=$30$). This
provided six visibilities, four closure phases, and four triple amplitudes
simultaneously in each of the eight 30-nm wide spectral channels
\citep{Monnier2004b, Monnier2006}. Using the same W1-W2-S2-E2 configuration of 
CHARA that was used for surface imaging of Altair \citep{Monnier2007}, we 
achieved adequate
({\it u,v}) coverage of each target for imaging. The longest baseline in this
configuration is~$251.3$~m, corresponding to a resolution of~$1.3$ mas at~$1.6
\mu$m. We secured 4 data blocks for RS~Per and 3 for T~Per, each data block
corresponding to a continuous observation of a target during about 20~minutes.

The data were reduced by the latest version of the MIRC pipeline written by
John Monnier (as of November 2012) and previously described in
\citet{Monnier2007}. The pipeline computes the squared visibilities using
Fourier Transforms, then averages them. The bispectrum is formed using the
phases and amplitudes of three baselines that form a closed triangle. For each
data block we use the best method available for amplitude calibration: for
T~Per the fluxes were estimated by the chopper method, and for RS~Per both the
chopper and DAQ method \citep{Monnier2008}. Our targets were observed along 
reference calibrators to correct for
the usual transfer function variations that occur during the night due to 
atmospheric and optical changes
in the beam path \citep{Perrin2003}. Our observations were typical of 
July-August weather, with transfer functions remaining very stable (less than 
0.2 drop in visibility during the nights).

The calibrators were modeled as uniform discs as indicated
in Table~\ref{tab:calibrators}. Note that $37$~And has recently been resolved
by MIRC as a binary ; however the flux ratio of the components is greater than
1:100 and thus this does not significantly impact our calibration.

As the brightness distributions of both targets is not expected to vary
significantly during our observing run (see Table~\ref{tab:calibrators} for the 
exact observing dates), all nights were combined into a single data
file for each target, resulting in data sets that total 419 power spectra and
264 bispectra for T~Per and 523 power spectra and 326 bispectra for RS~Per.
Systematic errors are taken into account by applying additive and
multiplicative errors on the data. All the following nominal values were 
determined based on the expertise of the MIRC group with MIRC 2007 data 
(Monnier, private comm.), and based on the in-depth study of $\upsilon$~And 
data 
acquired during the same nights \citep{Zhao2011}. Additive errors, that correct 
for biases at low fringe contrast were set to $2 \times 10^{-4}$ for squared 
visibilities and $10^{-5}$ for triple amplitudes. Multiplicative errors, that 
correct for the
uncertainties in the transfer function, were $20\%$ on squared visibilities
and $30\%$ on triple amplitudes. Based on the analysis of the closure phase
statistics made by \citet{Zhao2011}, an error floor of $1 \degree$ is chosen on 
closure phases. To account for the unreliability of closure estimation at low 
flux, each closure error is further increased by $30/S_\text{T3amp}^2$ degrees, 
where $S_\text{T3amp}$ is the signal-to-noise of the corresponding triple 
amplitude. Finally, to account for the difficulty of estimating rapidly varying 
closures, an error equal to 10\% of the closure gradient in the spectral domain 
is 
further added.

\section{Model-fitting}\label{sec:modeling}

\begin{figure}
\centering
\includegraphics[width=\linewidth]{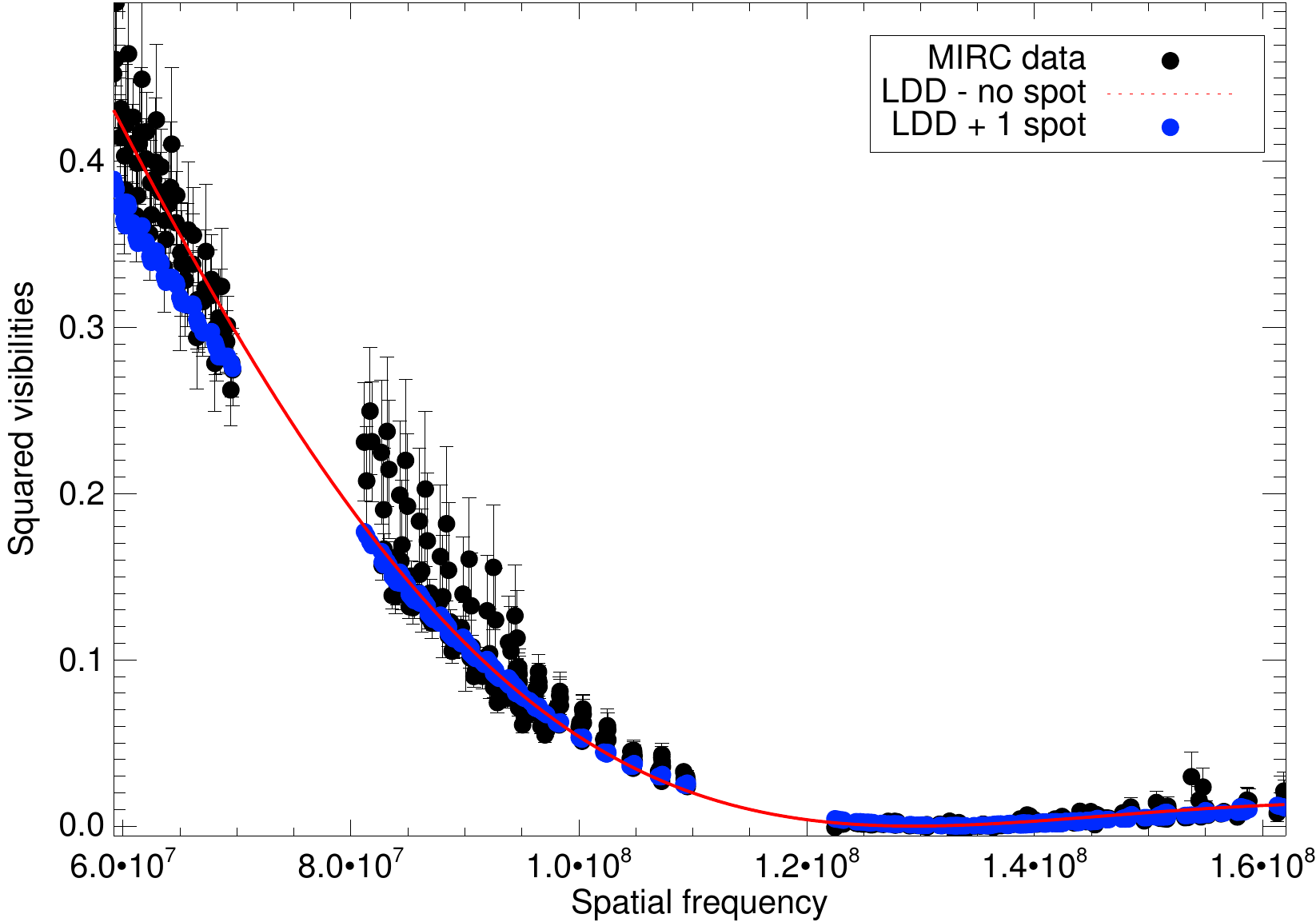}
\includegraphics[width=\linewidth]{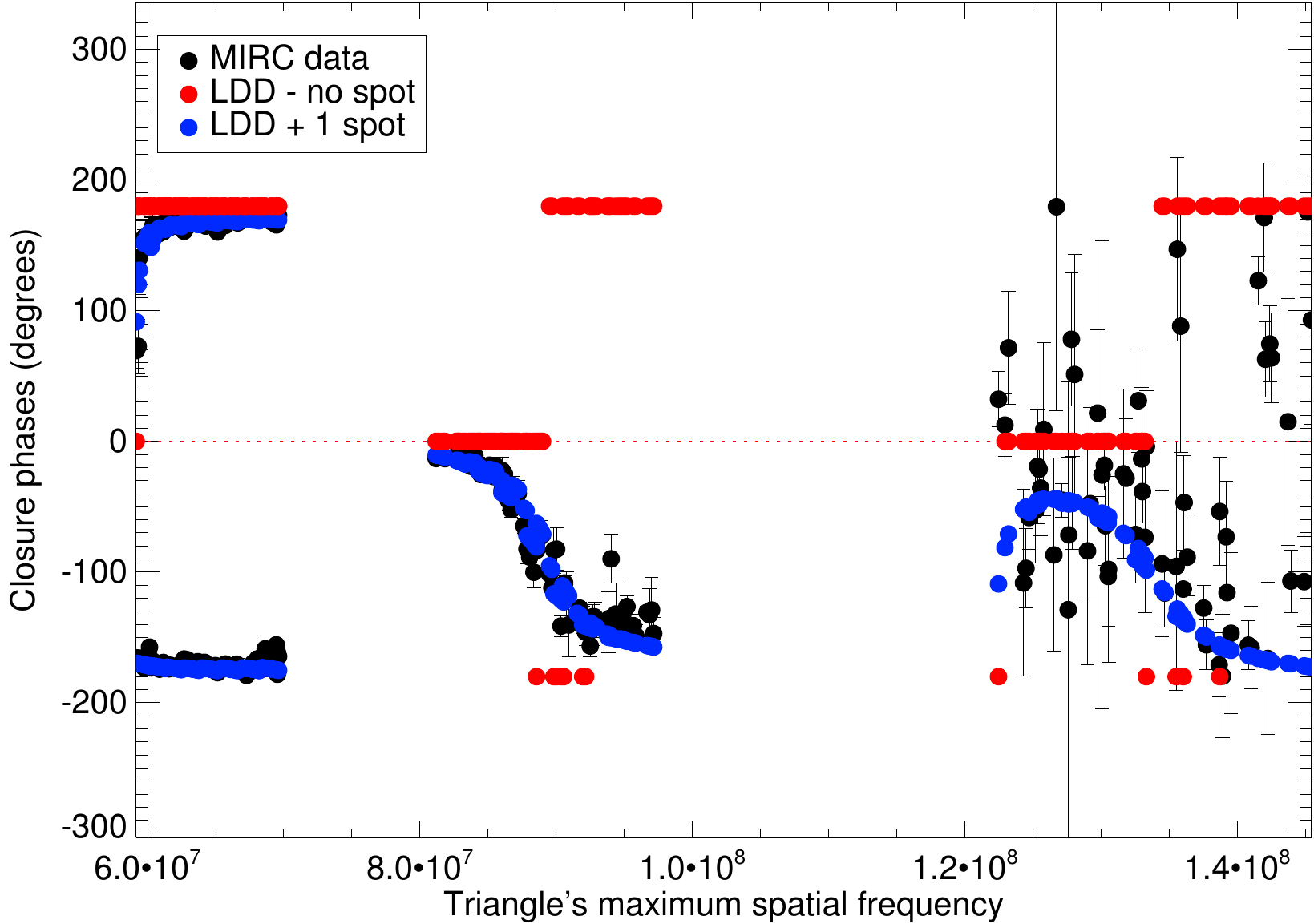}
\caption{Power spectrum and closure phase fits for T~Per. }
\label{fig:fits_tper}
\end{figure}

\begin{figure}
\centering
\includegraphics[width=\linewidth]{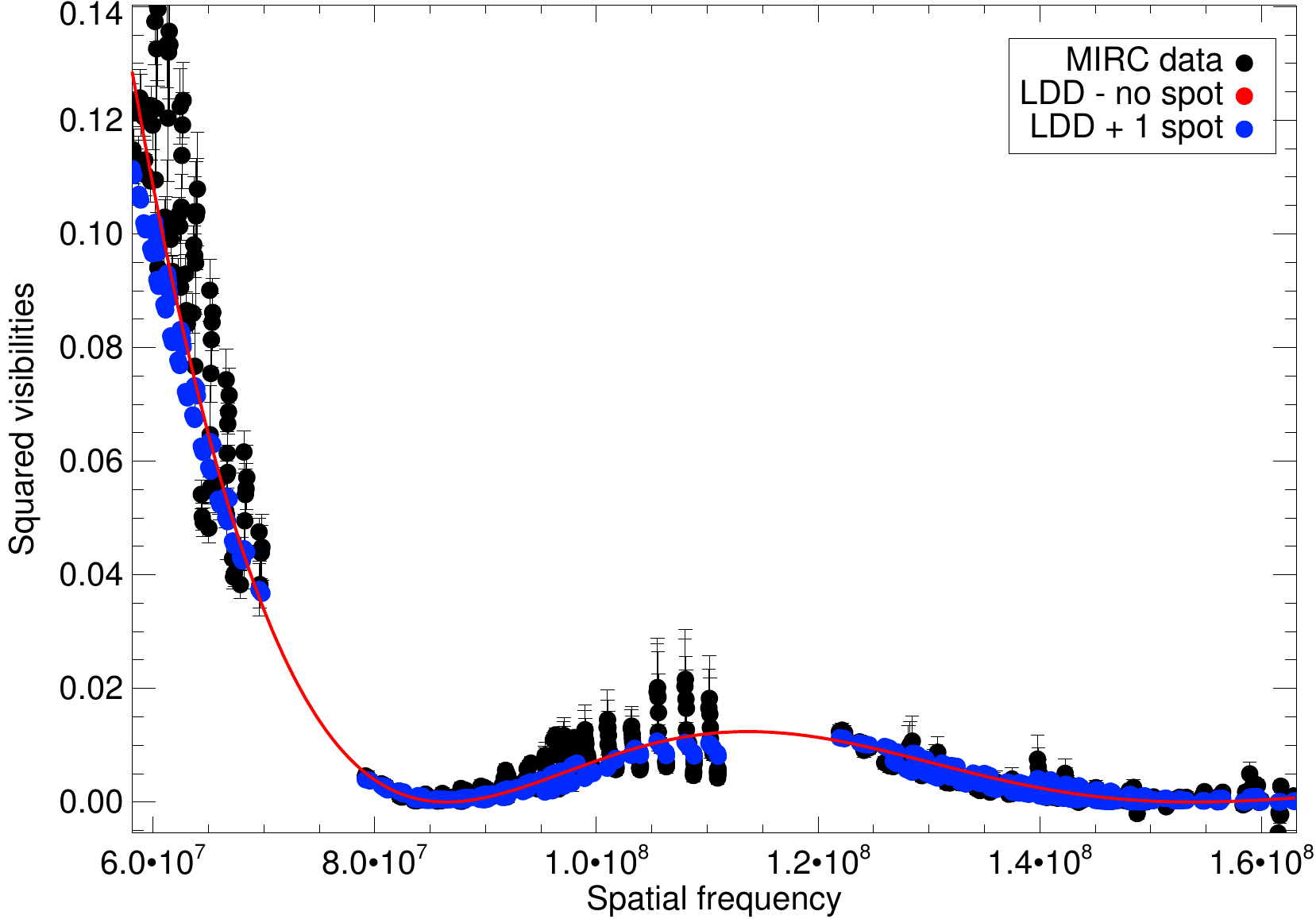}
\includegraphics[width=\linewidth]{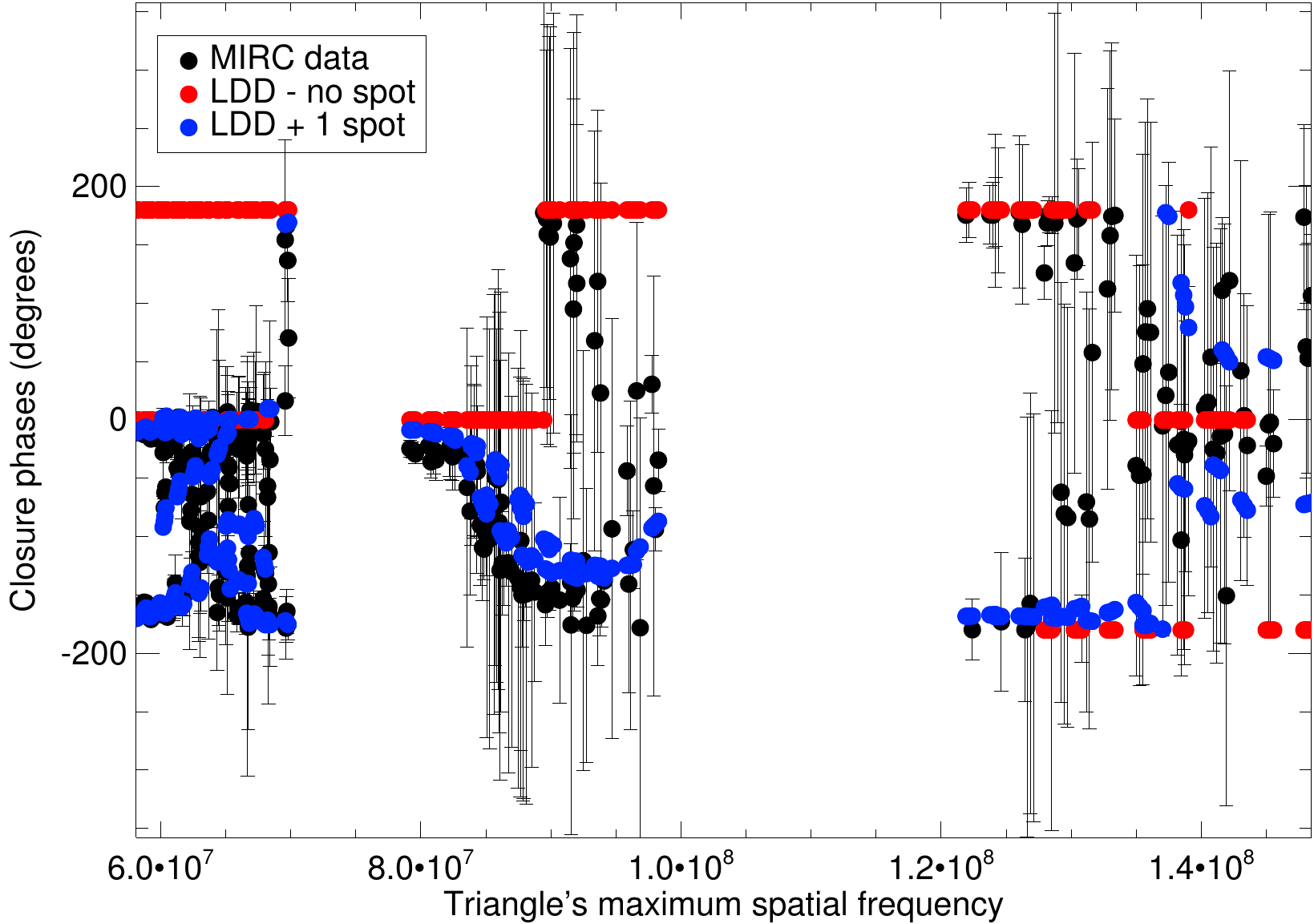}
\caption{Power spectrum and closure phase fits for RS~Per.}
\label{fig:fits_rsper}
\end{figure}

All the available square visibilities and closure phases are plotted on
Fig.~\ref{fig:fits_tper} and Fig.~\ref{fig:fits_rsper} as a function of their
spatial frequency. The power spectrum curves are typical of limb-darkened
stellar discs, while the closure phases clearly depart from zero, indicating
the presence of strong resolved asymmetries on the stellar surfaces. Based on
the previous interferometric results on M~supergiants in the literature, we
expect these to be due to the presence of spots.

\subsection{Spotless models: limb-darkening}\label{sec:ldd}

\begin{figure}
\centering
\includegraphics[width=0.5\linewidth]{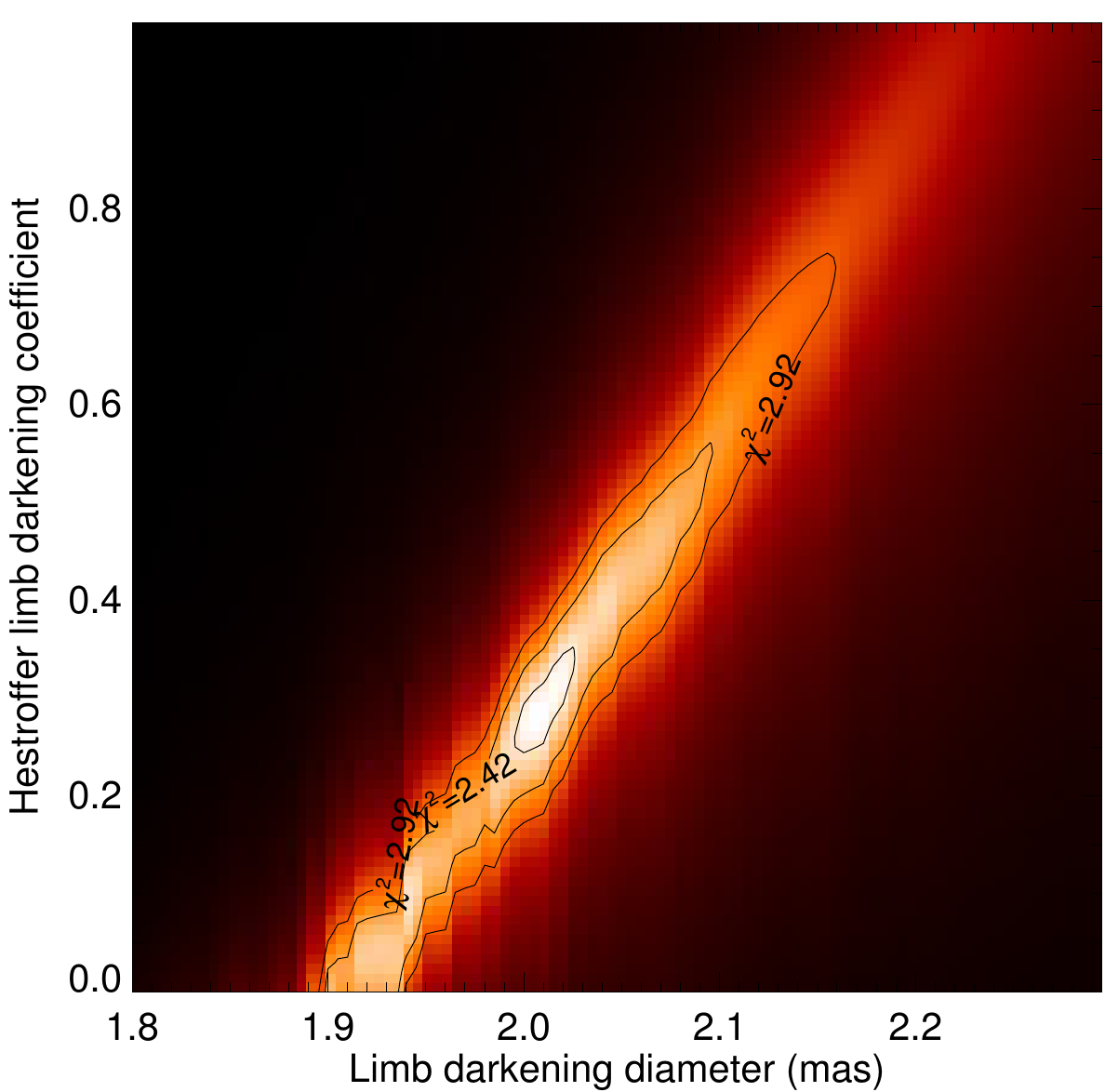}%
\includegraphics[width=0.5\linewidth]{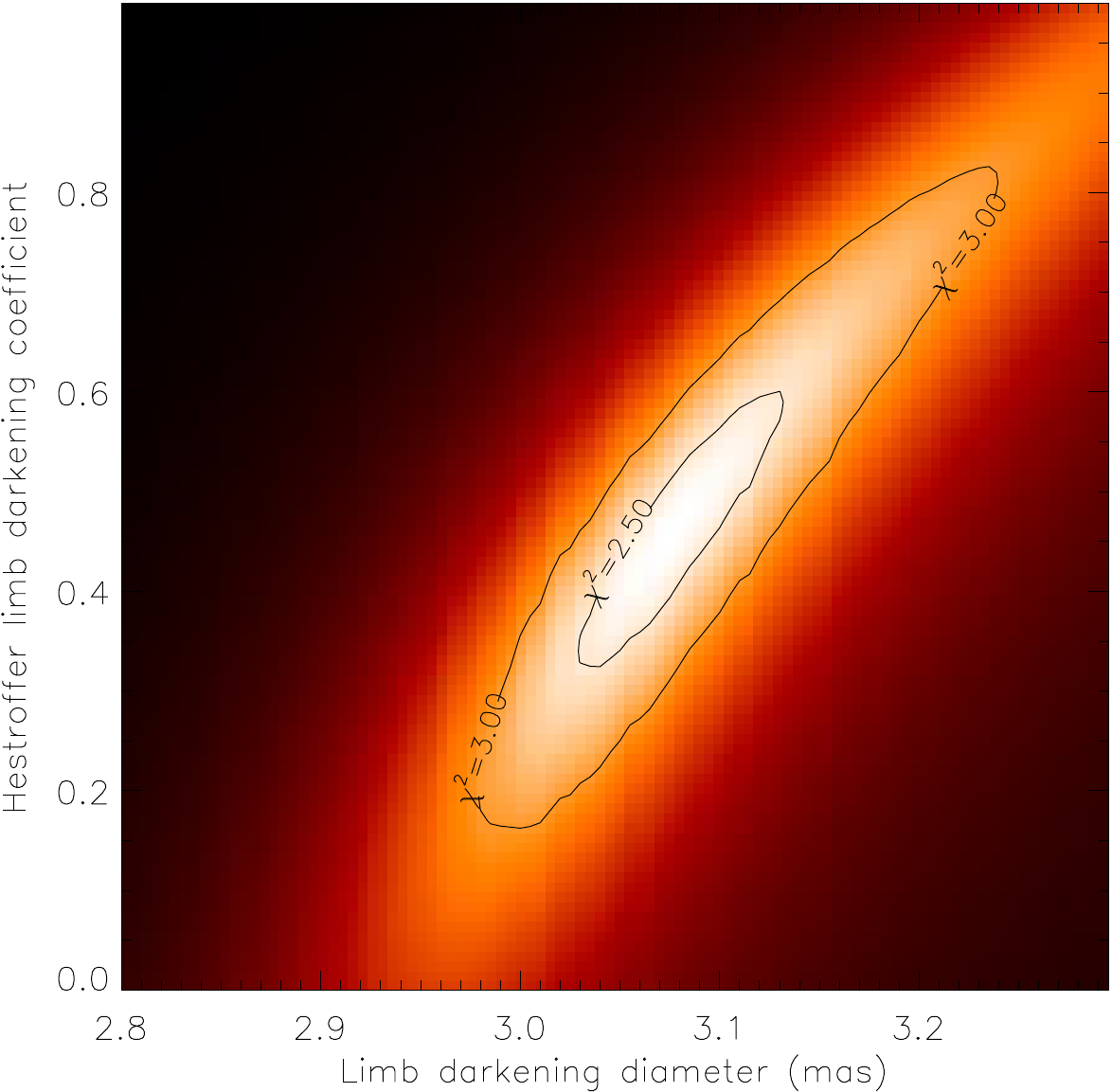}
\includegraphics[width=\linewidth]{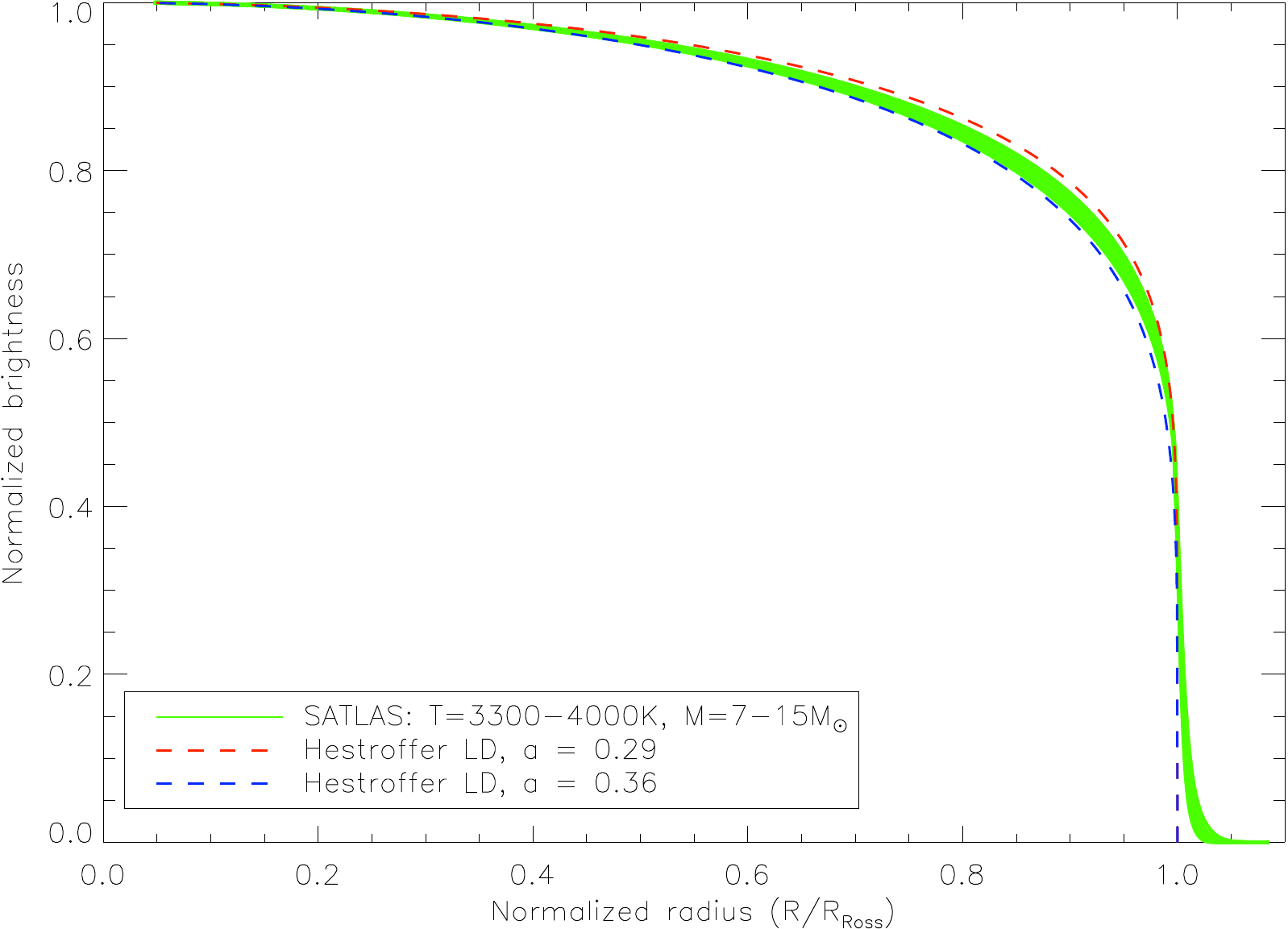}%
\caption{Constraints on the limb-darkening. Top panel: $\chi^2$ surface for the 
linearly limb-darkened discs of T~Per (left) and RS~Per (right) based on 
interferometric data. Bottom panel: SATLAS limb-darkening models 
for a range of temperature, gravity, and mass values compatible with previous 
observations of RS~Per and T~Per.}
\label{fig:ldd}
\end{figure}

Before attempting a spot search, we first sought to roughly characterize the
size and brightness distribution of the stellar discs. Our model-fitting code
FITNESS was used to fit several limb-darkening models (square root, quadratic,
power law) to the power spectra and triple amplitudes. FITNESS is a 
straightforward combination of grid search to identify the
global $\chi^2$ minima and Levenberg-Marquardt to refine the parameters. The
best fits were obtained for the linear law and the Hestroffer power law
\citep{Hestroffer1997}, but they both show the existence of a strong
covariance between the limb-darkened angular diameter and the limb-darkening
coefficient. The issue is illustrated on Figure~\ref{fig:ldd} (top), where the
reduced $\chi^2$ surface is plotted as a function of both parameters. For both
targets, the problem is mainly due to the lack of high signal-to-noise data on
the first visibility lobes. To increase the precision of the fit on the
angular diameter, the limb-darkening coefficients have to be constrained.

\citet{Haubois2009} reported successfully fitting the visibility curve of
$\alpha$~Ori (M2 type, $T_{\text{eff}} \simeq 3600$~K) with a linear
coefficient $\alpha = 0.43 \pm 0.03$ (roughly corresponding to a Hestroffer law 
with coefficient~$0.3$--$0.4$). Beyond this empirical result, it seems
non-obvious whether conventional plane-parallel 1D radiative codes such as
ATLAS \citep{Kurucz1992, Castelli2004} or MARCS \citep{Gustafsson1975} may be
reliably predict the intensity profiles of RSGS. RSGs are notoriously
difficult to model: their atmospheres are very extended, which invalidates the
assumption of plane-parallel geometry, and their very cool temperatures
require an advanced treatment of molecular opacities. However more recent
codes that assume spherical geometry such as MARCS-spherical
\citep{Gustafsson2008}, SATLAS \citep{Lester2008} and PHOENIX
\citep{Hauschildt1999} have demonstrated successful results on comparably cool
M~giants or supergiants \citep{Wittkowski2004, Wittkowski2006, Wittkowski2012}.

We used the latest SATLAS code (with improved ODF treatment and fixed H2O
lines) to weakly constrain the limb-darkening in H~band for both RSGs, with 
parameters based around the values found in \citet{Gonzalez2000} and 
\citet{Slesnick2002}: temperatures ranging from 3100~K to 4000~K (steps of 
100K), $\log g = -0.5$ to 
$0.5$ and a fixed metallicity $[\text{Fe/Z}]=-0.5$ \citep{Gonzalez2000}. 
Figure~\ref{fig:ldd} (bottom) presents the
results of these simulations as a band of possible brightness distributions. 
The intensity profiles
are weakly dependent on the temperature and are mostly determined by the mass
and surface gravity. They are characteristics of spherical codes, showing a
sudden drop of intensity at the Rosseland radius, where the Rosseland mean 
opacity equals unity and where most photons escape the atmosphere. As shown on 
Figure~\ref{fig:ldd}, the brightness distributions can adequately be bounded by 
two Hestroffer laws with coefficients~$0.29$ and~$0.36$. Injecting this prior 
into the fit, we found the Hestroffer limb-darkening coefficients to be $\alpha 
= 0.32 \pm 0.2$ for T~Per and $\alpha = 0.34 \pm 0.2$ for RS~Per. Assuming that 
both RSGs have similar masses, as the angular diameter of T~Per is smaller, we 
expect a stronger gravity at its surface, which indeed corresponds to a lower 
limb-darkening coefficient.

Taking into account all statistical errors due to the 
visibility measurements and the calibration via data bootstrapping, we also 
obtained the following
limb-darkened diameters: for T~Per: $\theta_{\rm LD} = 2.01\pm 0.03$~mas, with
$\chi^2=1.92$ at the nominal values. For RS~Per, $\theta_{\rm LD} =
3.05\pm 0.05$~mas and with $\chi^2=2.37$.  Because non-zero closure
phases cannot be fitted by a limb-darkening model, the ``full'' $\chi^2$ --
including the closure phase data -- are larger, $\chi^2 = 7.8$ for RS~Per and
$\chi^2 = 9.4$ for T~Per, unambiguously indicating the presence of significant
asymmetries on the stellar surfaces.

\subsection{Models with one or two spots}

\begin{figure}
\centering
\includegraphics[width=0.5\linewidth]{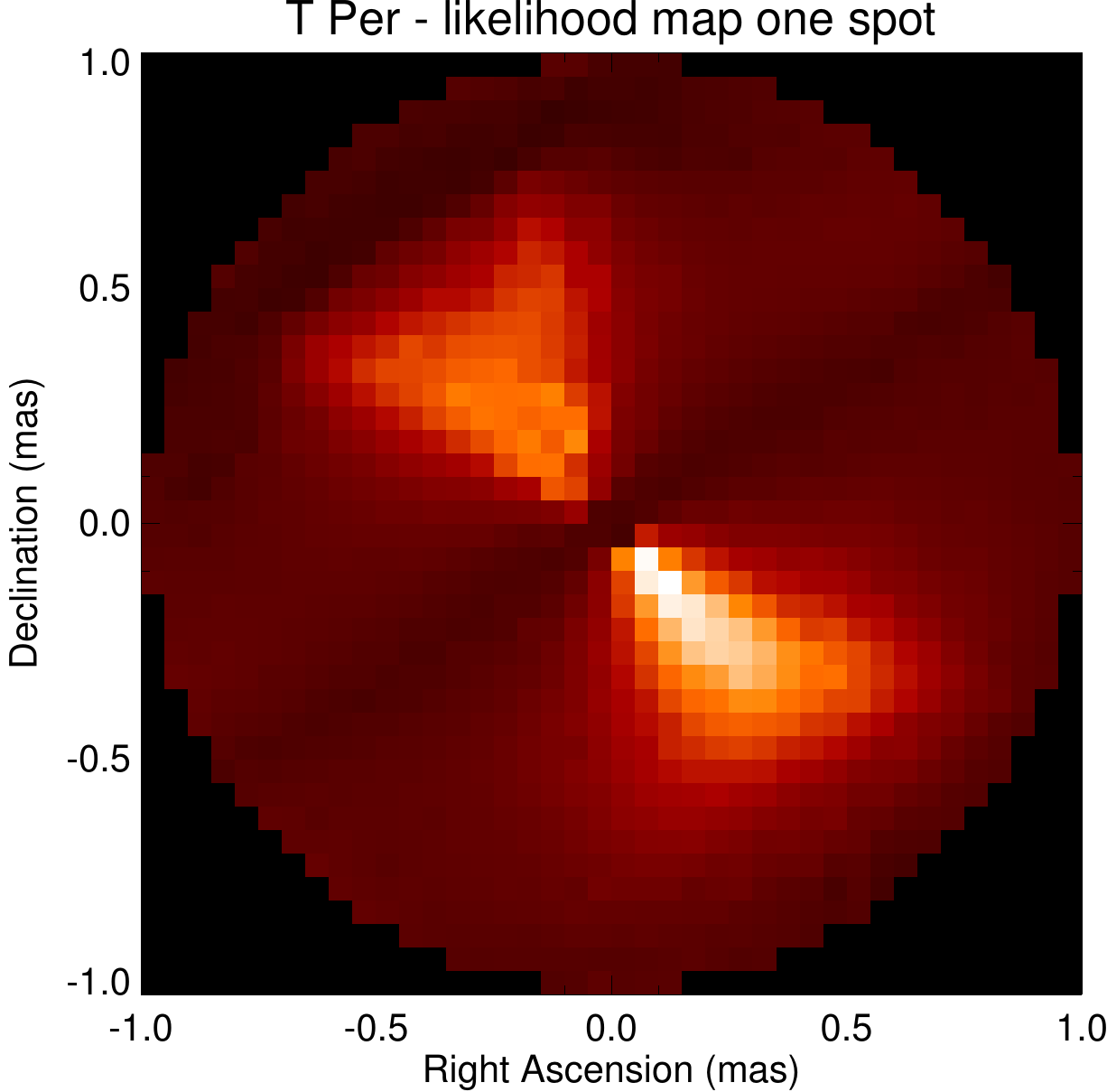}%
\includegraphics[width=0.5\linewidth]{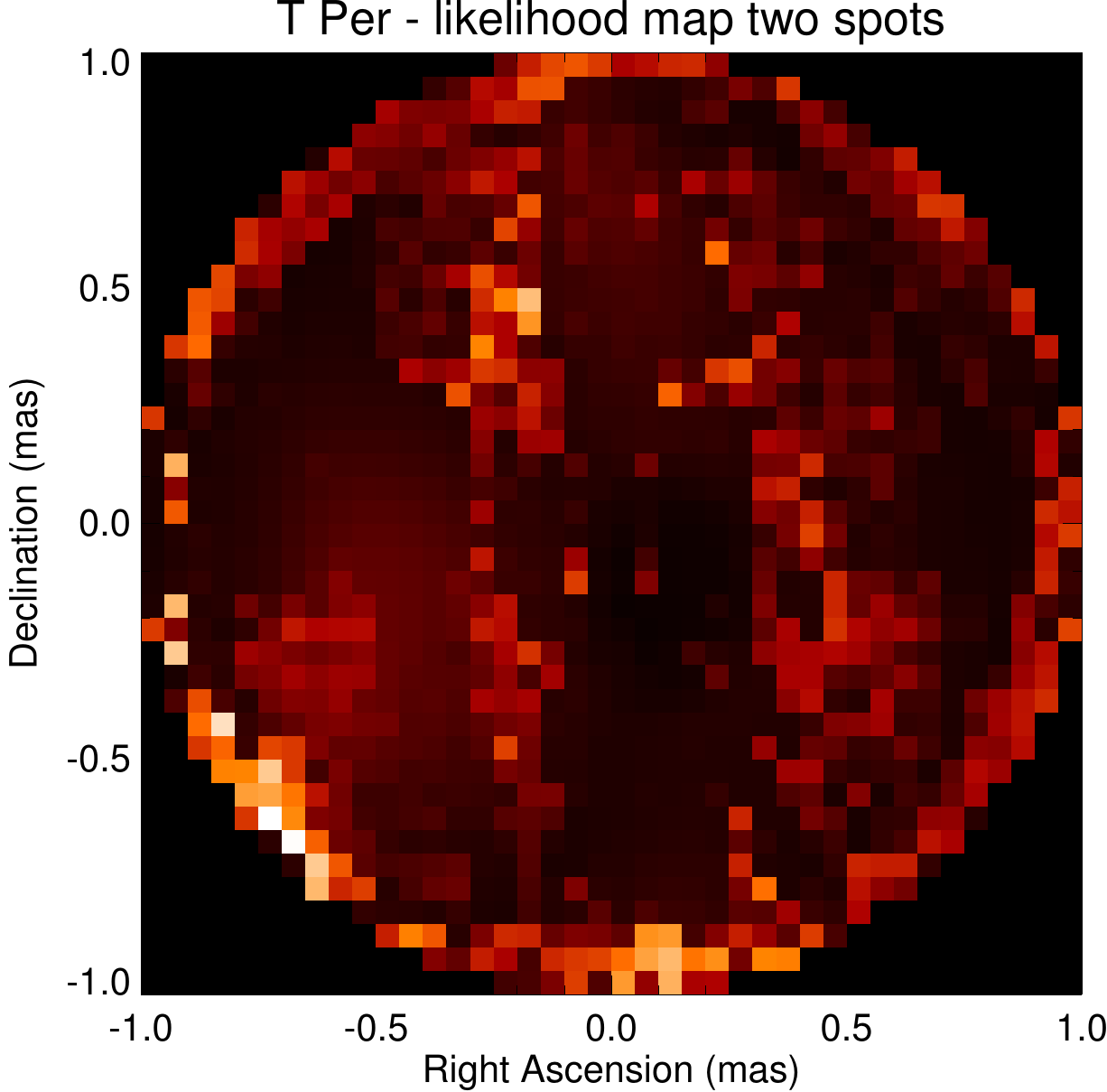}
\includegraphics[width=0.33\linewidth]{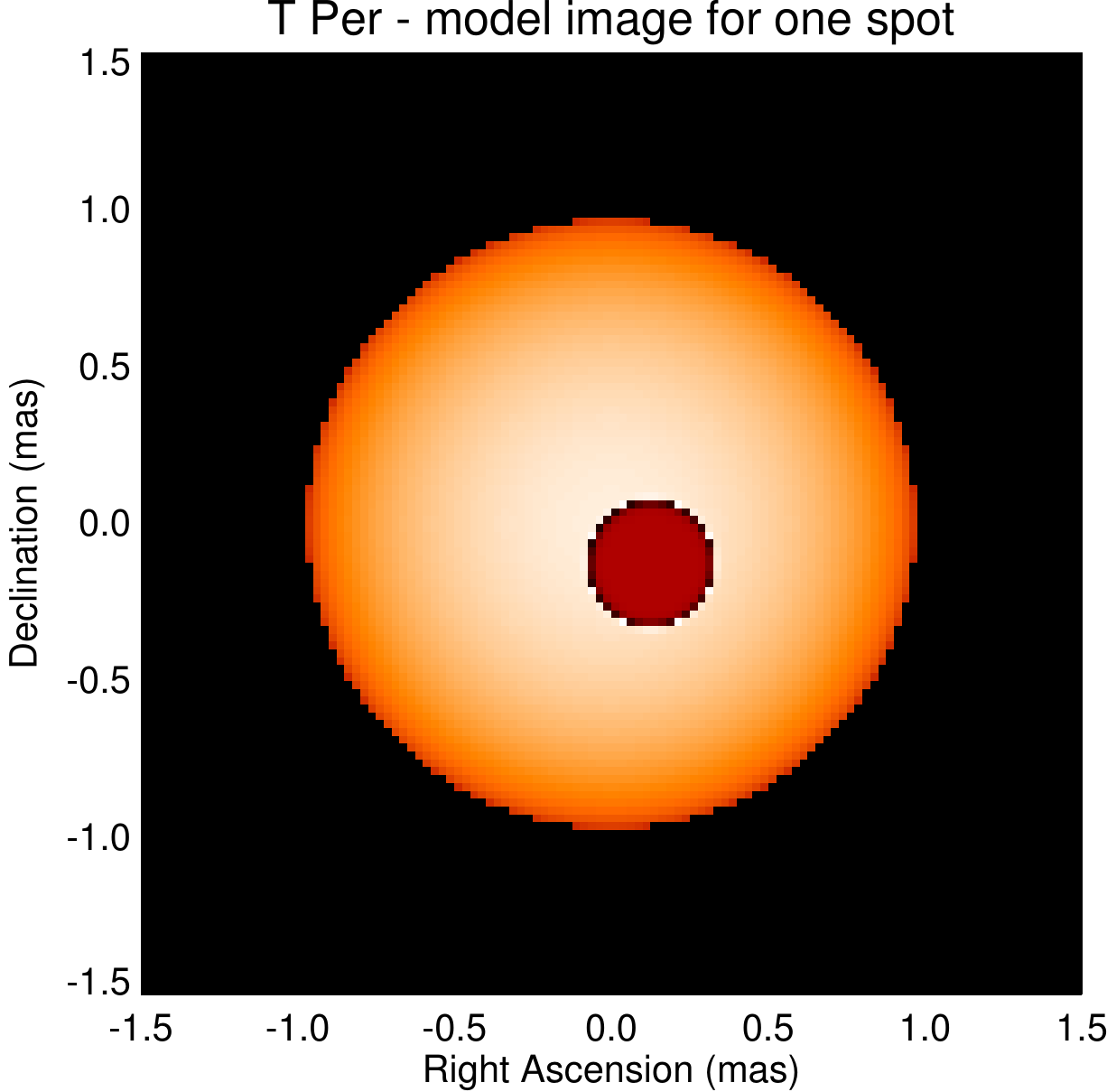}%
\includegraphics[width=0.33\linewidth]{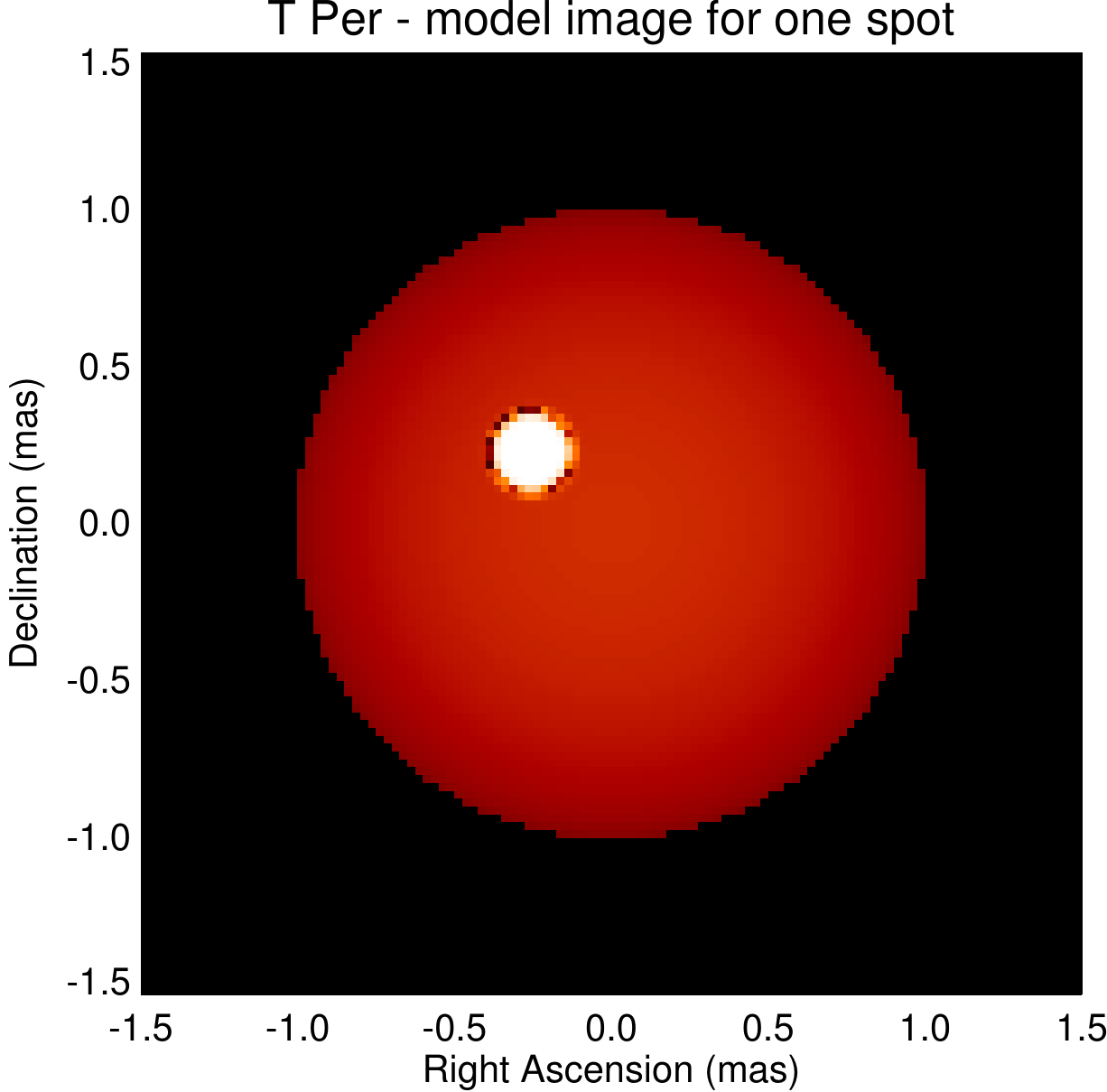}%
\includegraphics[width=0.33\linewidth]{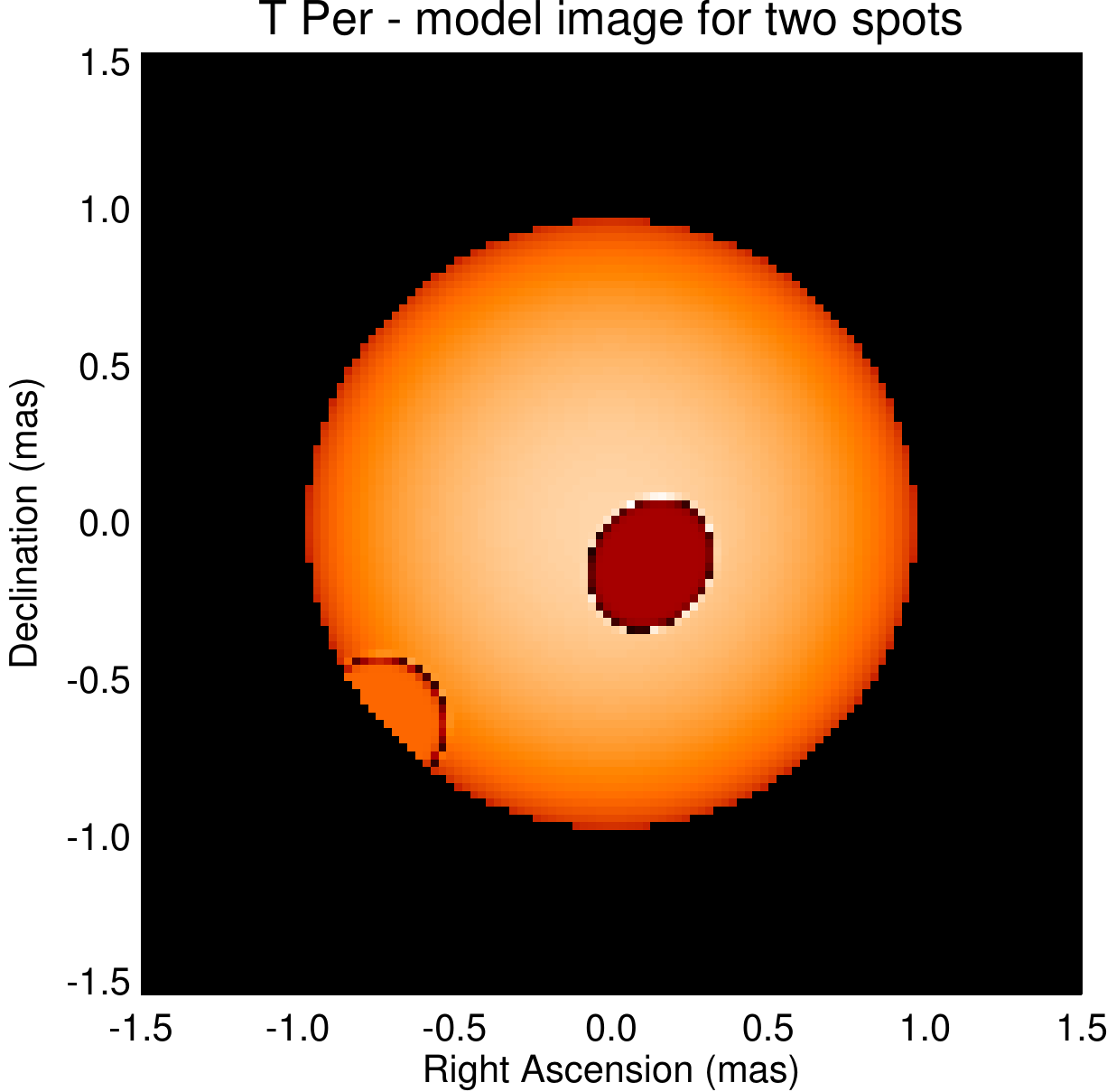}
\caption{Simple modeling of T~Per: likelihood maps for one- and two-spot models 
(top); corresponding best fitting images for a dark spot, bright spot and two 
spots (bottom).}\label{fig:tper_spots}
\end{figure}

\begin{figure}
\centering
\includegraphics[width=0.5\linewidth]{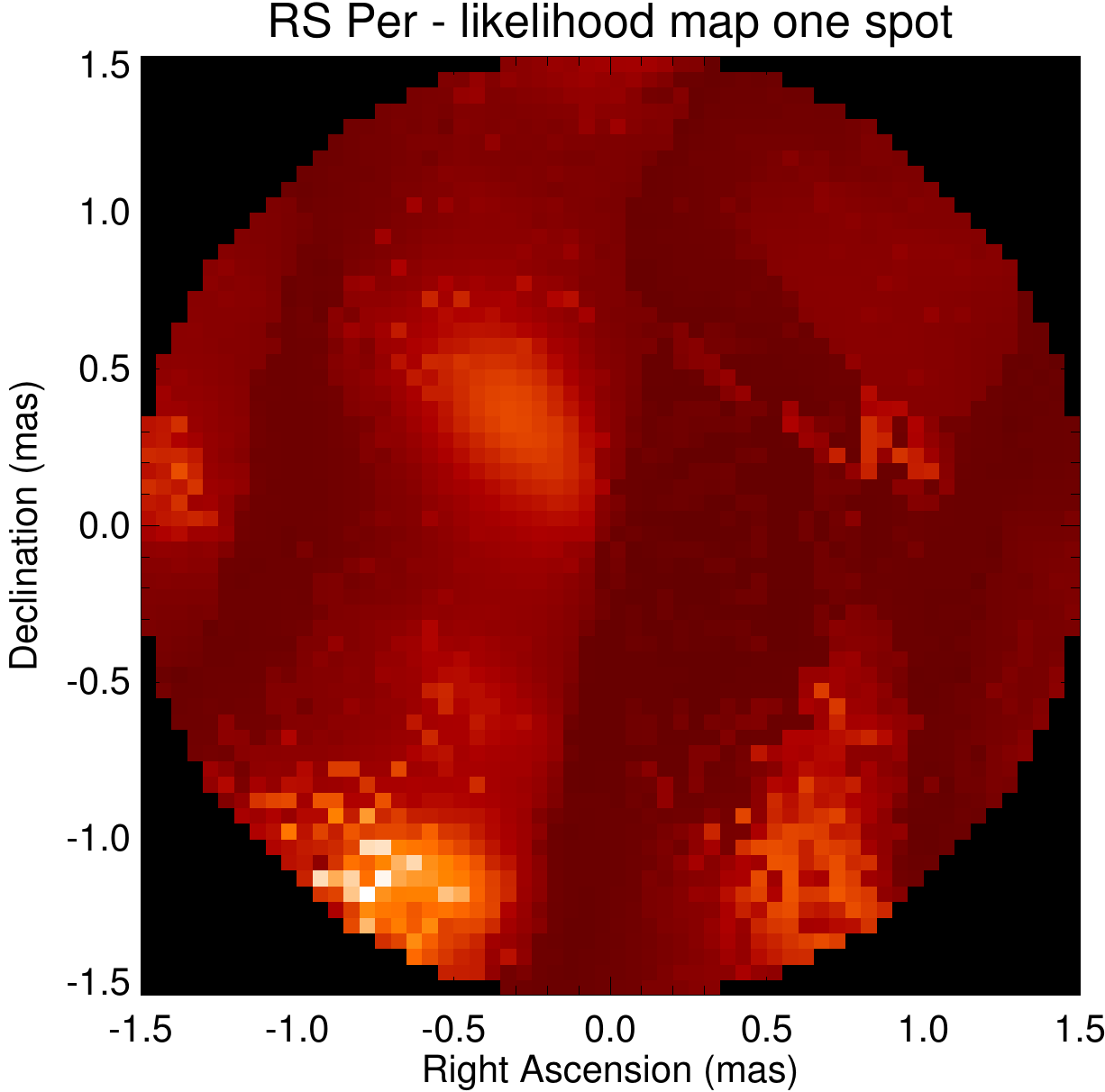}%
\includegraphics[width=0.5\linewidth]{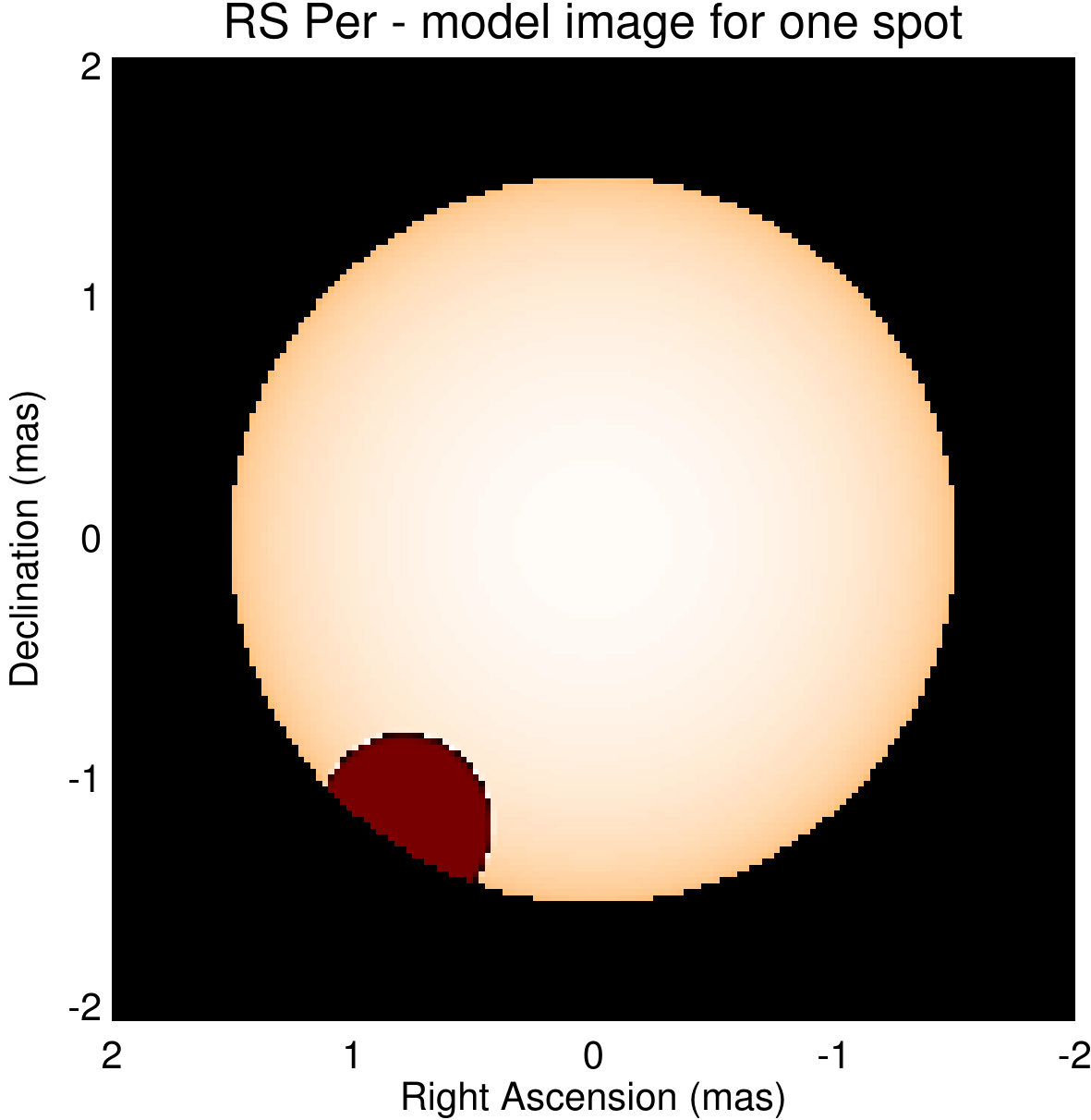}
\caption{Simple modeling of RS~Per: likelihood map of the spot location (left) 
and best fitting image with a dark spot (right).}\label{fig:rsper_spots}
\end{figure}

\begin{deluxetable}{lcccc}
\centering
\tabletypesize{\scriptsize}
\tablecaption{Model-fitting Results for T~Per (angles given East of North).}
\tablewidth{0pt}
\tablehead{ Fit results & \colhead{No spot} & \colhead{Dark spot} & 
\colhead{Bright spot} & \colhead{Two spots} }
\startdata
$\chi^2$ & $9.4$ & 2.24 & 2.64 & 1.95 \\
$\log Z (\pm 0.02)$ & $-0.35$ & $0.87$ & $1.23$ & $0.20$ \\
$\theta_{\star}$ & 2.01 & 2.02 & 2.02 & 2.02 \\ 
$f_{\text{spot}}$ & $\ldots$ & 4 \% & 5\% & 3\%, 4\% \\
$(r_{\text{spot}}, \theta_{\text{spot}})$ & $\ldots$ & $0.21, -135 \degree$ & 
$0.22, 
134 \degree$ & $0.22, 134 \degree$ \\
& & & & $0.73, 238 \degree$ \\
\enddata 
\label{tab:tper_fits}
\end{deluxetable}

\begin{deluxetable}{lcc}
\centering
\tabletypesize{\scriptsize}
\tablecaption{Model-fitting Results for RS~Per (angles given East of North).}
\tablewidth{0pt}
\tablehead{ Fit results & \colhead{No spot} & \colhead{Single dark spot} }
\startdata
$\chi^2$ & $7.8$ & 1.2 \\
$\log Z (\pm 0.05)$ & $-0.35$ & $0.32$ \\
$\theta_{\star}$(mas) & 3.05 & 3.06 \\ 
$f_{\text{spot}}$ & $\ldots$ & 4 \% \\
$r_{\text{spot}}$(mas), $\theta_{\text{spot}}$ & $\ldots$ & $1.43, 147 \degree$ 
\\
\enddata 
\label{tab:rsper_fits}
\end{deluxetable}

The most economical assumption to explain the closure phases is the presence
of spots on the stellar surface. The formation of complex granulations is
expected on the surface \citep{Freytag2002, Chiavassa2011}, and at the
resolution and dynamic contrast provided by interferometry, these appear as
compact spots. Here we attempted to search only for the most prominent
features, with the assumption that the stellar surface could be described as a
limb-darkened disc with a limited number of these spots. Previous studies on
$\alpha$~Ori demonstrated that spots may be modeled well by Gaussians or
uniform ellipses \citep{Young2000, Haubois2009}. We also chose
to use ellipses, so that each spot is modeled by six parameters: its
coordinates on the stellar surface, its size, its flux contribution, its
orientation angle and its ellipticity. Our priors on these parameters were
flat. During model-fitting, spots were not constrained to be brighter than the
stellar surface (hot) and thus dark (cold) spots were not ruled out. Spots 
lying on the circumference of the stellar disc were also searched for.

Fitting spots is a difficult numerical problem due to the properties of the
$\chi^2$. First, while the bispectrum probability density is generally 
approximated by a convex normal distribution \citep{Meimon2005, Thiebaut2008}, 
this approximation breaks down for small triple amplitudes, e.g. for very 
resolved targets such as these RSGs. Therefore we revert to use separate 
$\chi^2$~expressions for closures and triple amplitudes. The likelihood 
expression for the closure phases may then be chosen based $2\pi$-wrapped 
normal distribution \citep{Haniff1991} or the von~Mises distribution, and in 
general is non-convex. Moreover, because the phase information is only partially
retrieved from closure phases, the $\chi^2$ is multimodal,
i.e. there exist local minima into which minimizers can easily get trapped
\citep{Meimon2008}. The $\chi^2$-minimization strategy has then to take both
non-convexity and multimodality into account. Due to the relatively large
number of parameters (eight to fourteen: two for the stellar disc description,
plus six per spot), our strategy consisted in a grid search on the positions
of the spots, with a three-step Levenberg-Marquardt minimization at each given
position. During the first step, only the stellar disc parameters (diameter
and Hestroffer coefficient) and the spot flux are allowed to vary. During the
second step, the shape of the spots (size, aspect ratio, orientation) is
optimized, then in the third step consists all the parameters become free to
settle.

Figure~\ref{fig:fits_tper} and Figure~\ref{fig:fits_rsper} show that overall
the fits provided by our models are not very satisfactory on the short
baselines, where our models underestimated the visibilities. This is due to
our naive assumption of a linearly limb-darkened disc model, compared to more
refined models (e.g. SATLAS, Figure~\ref{fig:ldd}), and possibly to the
presence of circumstellar material further obscuring the stellar discs at
their edges. Note that model-independent image reconstruction, carried out in
Section~\ref{sec:image_reconstruction}, also shows evidence for strong
darkening at the periphery. However these considerations do not significantly
affect our spot analysis, relying on medium and long baseline
visibilities. Figure~\ref{fig:tper_spots} and Figure~\ref{fig:rsper_spots}
presents the $\chi^2$ surface as a function of the spot position, as well as
the corresponding best one-spot and two-spot models for T~Per and RS~Per. In
all cases, we found that the total flux contribution of the spots represent
roughly $\sim 4\%$ of the stellar flux. The reduced $\chi^2$ and parameters
for these models are given in Table~\ref{tab:tper_fits} and
\ref{tab:rsper_fits}, with error bars derived using the classic bootstrapping
technique. For T~Per, our results imply the presence of a spot on along the
diagonal NW-SE. Based solely on the $\chi^2$ metrics, a dark spot in the SE
quadrant ($\chi^2 = 2.24$), or a bright spot on the NW quadrant seem equally
probable ($\chi^2 = 2.35$). There is also a slight decrease of $\chi^2$
($\chi^2 = 1.9$) when attempting to fit an additional spot to the dark spot
model. For RS~Per, the results are clearer, with evidence of a single dark
spot in the SW. In particular no solution involving any bright spot could be
found.

\subsection{Bayesian spot model selection}

In general the reduced-$\chi^2$ metric is ill-adapted to truly assess the
relative probabilities of models \citep{Marshall2006}. The $\chi^2$ decrease
that results from the addition of a new set of spot parameters can be due to
modeling a real spot or simply to over-fitting, with emergence of artifacts
due to imperfect ({\it u,v}) coverage. Here we present a general framework to
treat the problem of fitting spots, based on Bayesian model
selection. Model-fitting consists in estimating the most probable model
coefficients $\V{C}=\{ c_1, c_2, \ldots c_n\}$ of a model $\V{M}_i$. To do so
we actually maximize the posterior probability $p(\V{C}|\V{D},\V{M}_i)$:
\begin{equation}
 p(\V{C}| \V{D}, \V{M}_i)= \frac{p(\V{D}|\V{C},\V{M}_i) 
p(\V{C}|\V{M}_i)}{p(\V{D}|\V{M_i})} , \label{eq:bayes}
\end{equation}
where $p(\V{D}|\V{C},\V{M}_i)$ is the likelihood of the model coefficients,
$p(\V{C}|\V{M}_i)$ is the prior probability of the coefficients, and
$p(\V{D}|\V{M_i})$ is the marginal likelihood, also often called ``Bayesian
evidence'' in astronomy. The evidence appears in Eq.~\ref{eq:bayes} as the
denominator, its role being to normalize the posterior probability.  For a
given model $\V{M_i}$, the evidence is constant. To determine the $\V{C}$
coefficients it is then sufficient to maximizing the denominator only: i.e the
likelihood (i.e. $\chi^2$) under prior constraints (mostly physical
constraints such as positivity for the stellar diameters, spots within the
stellar discs).  However when comparing two spot models, the ratio of their
evidence is to be considered. The ratio of the probabilities of two models
$\V{M}_1$ and $\V{M}_2$ given the data can be expressed as:
\begin{equation}
\frac{p(\V{M_1}|\V{D})}{p(\V{M}_2|\V{D})}= \frac{p(\V{M}_1)}{p(\V{M}_2)}
\frac{p(\V{D}|\V{M_1})}{p(\V{D}|\V{M_2})} \label{eq:model_selection}
\end{equation}
where $p(\V{M}_i)$ is the {\it a priori} probability of model $\V{M}_i$. As we
have no specific preference for a model in the absence of data,
$p(\V{M}_1)/p(\V{M}_2) = 1$, and therefore the most probable model corresponds
to the model with the largest evidence. For a given data set, the evidence
$Z(\V{M_i}) = p(\V{D}|\V{M_i})$ for model $\V{M_i}$ is defined as the
marginalized likelihood:
\begin{equation}
Z(\V{M_i}) = \int\limits_{C_1}\ldots\int\limits_{C_n} p(\V{D}|\V{C},\V{M}) 
p(\V{C}|\V{M})
dC_1 \ldots dC_n. \label{eq:evidence_marginalization}
\end{equation}
Computing the evidence and its associated error bar with good precision
requires the exploration and integration of the posterior probability by
specialized Monte-Carlo Markov Chain algorithms. Our model-fitting code
FITNESS uses the MultiNest library \citep{Feroz2008, Feroz2009} to compute the
logarithm of the evidence $\log Z$ with the Nested Sampling algorithm
\citep{Skilling2006}. Because the ratio of evidence intervenes in
Eq.~\ref{eq:model_selection}, differences of $\log Z$ encode the relative
model probabilities and can be interpreted (with caution) using the Jeffrey's
scale \citep{Kass1995}. Contrary to the reduced-$\chi^2$, $\log Z$ does not
directly take into account the raw number of parameters, but it is based on
their actual relevance to the fitting process. A good model has the minimal
number of parameters required to explain the data (Occam's razor), which
corresponds to a high $\log Z$. Bad models may be less predictive, or may be
too generic due to the overabundance of parameters, and the are characterized
by low $\log Z$. We give the $\log Z$ for all our models in
Table~\ref{tab:tper_fits} and Table~\ref{tab:rsper_fits}.  For T~Per, $\log Z$
points overwhelmingly in favor of the single bright spot model. Despite having
better $\chi^2$, the single dark spot and the two-spot models are found to be
much less probable. For RS~Per, a comparison of the $\log Z$ for the spotted
and non-spotted models indicate that the dark spot in the SW is probably real,
though with a low confidence index. Both these results will be confirmed by
image reconstruction in Section~\ref{sec:image_reconstruction}.

\subsection{Effective surface temperatures}\label{sec:temperatures}

The effective temperature of a star $T_{\rm eff}$ and its bolometric flux
$F_{\rm bol}$ follow the Stefan-Boltzmann law, $F_{\rm bol} = \sigma T_{\rm
  eff}^4$. The measured bolometric flux on Earth $f_{\rm bol}$ is weaker by a
factor $\theta^2/4$, where $\theta$ is the angular diameter of the star. The
effective temperature is then given by:
\begin{equation}
T_{\rm eff}= \left(\frac{4 f_{\rm bol}}{\sigma_B 
\theta^2}\right)^{\frac{1}{4}},\label{eq:temperature}
\end{equation}
where $\sigma_B$ is the Stefan-Boltzmann constant. As recommended by 
\citet{Scholz1987}, $\theta$ is chosen to be the Rosseland angular diameter. 
Considering the relatively low signal-to-noise of the data, we assume here the 
Rosseland diameter to be equal 
to the limb-darkened diameter $\theta_{\text{LD}}$ fitted in~\ref{sec:modeling}.

\begin{figure}
\centering
\includegraphics[width=\linewidth]{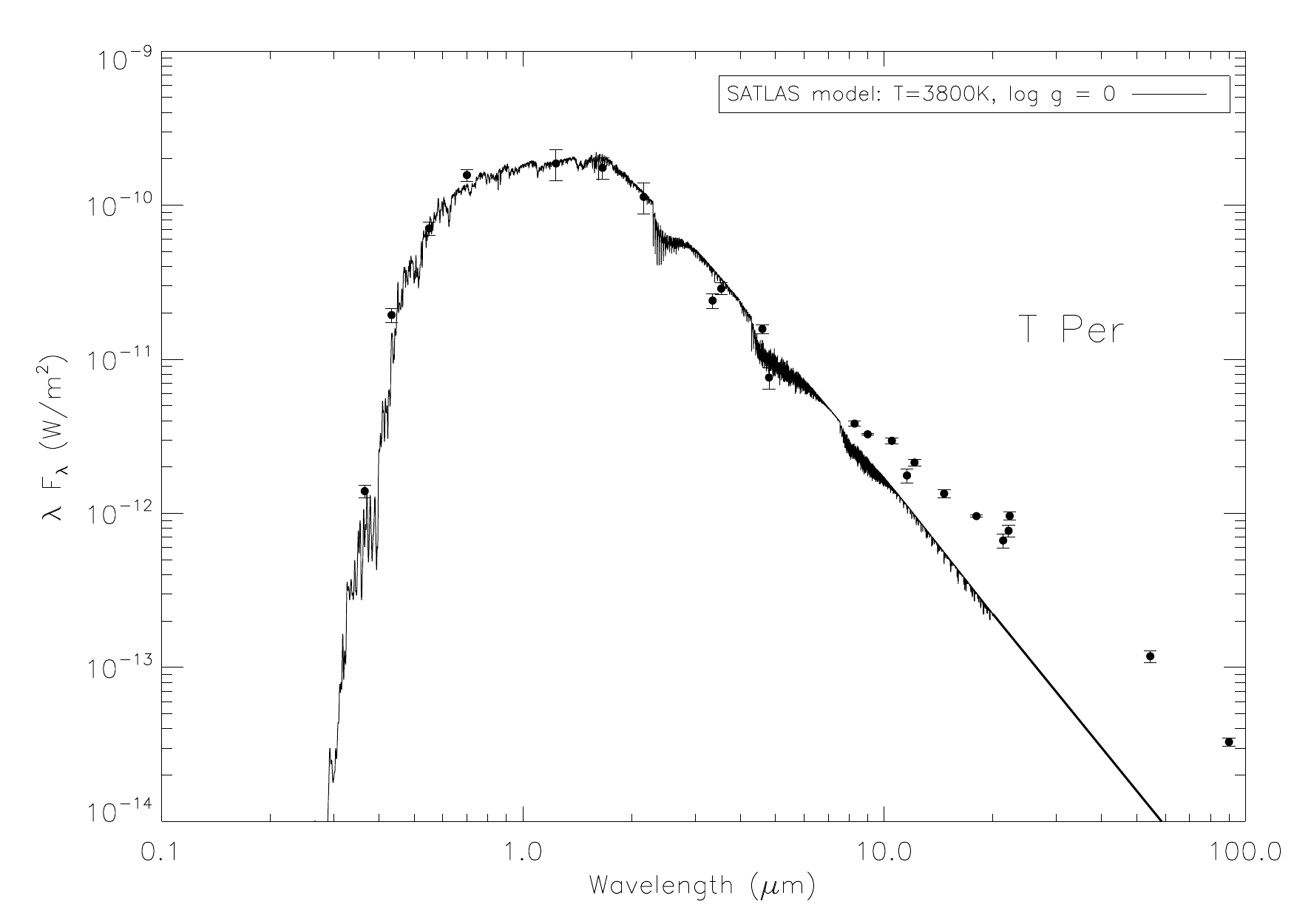}\\
\includegraphics[width=\linewidth]{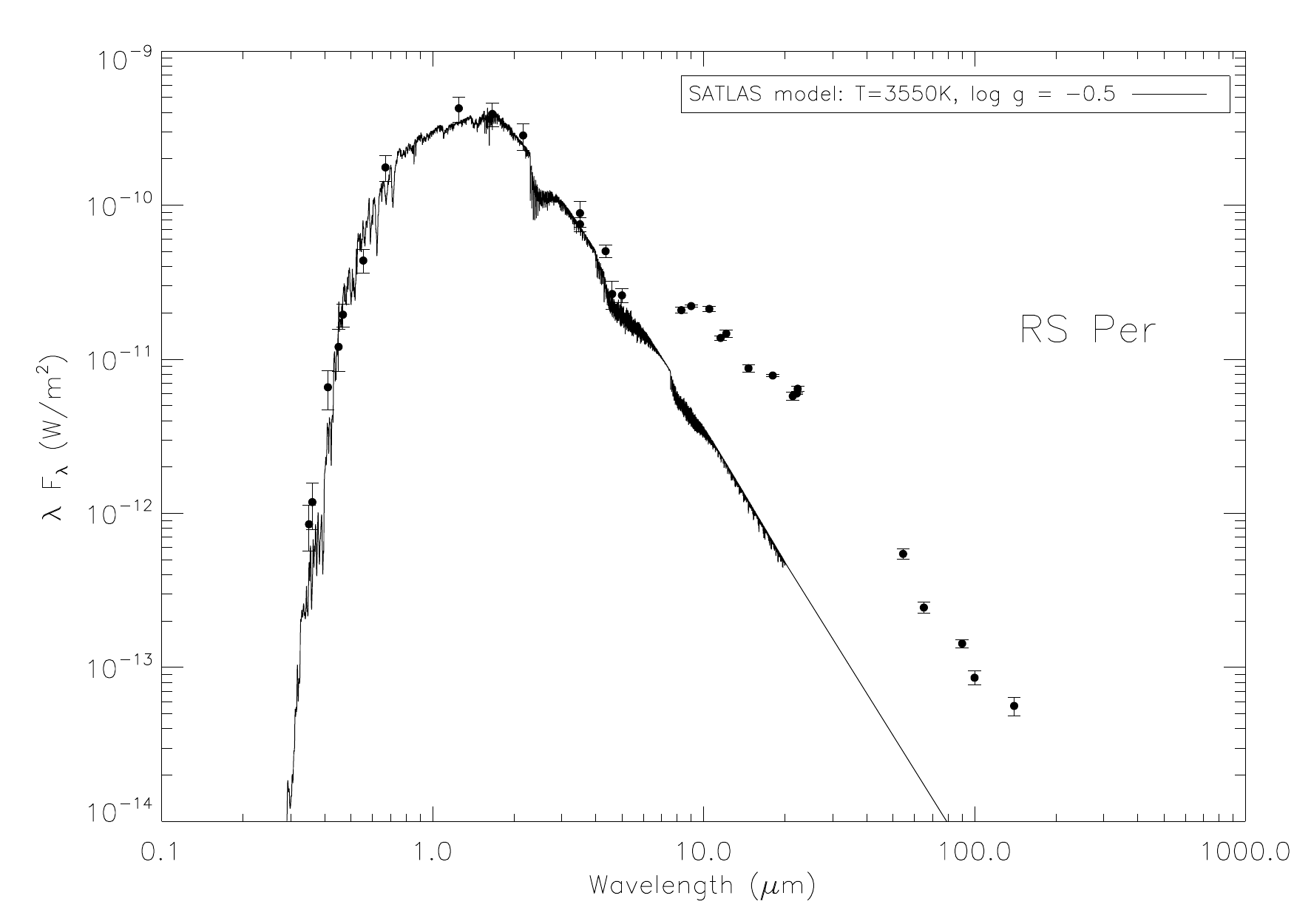}%
\caption{Spectral energy distributions of T~Per (top) and RS~Per (bottom). The 
plain
lines show the dereddened SATLAS fit to the optical and near-IR parts of the 
spectra.}\label{fig:seds}
\end{figure}

To compute $f_{\rm bol}$ for our targets, we derived the spectral energy
distributions (SEtDs) from public catalog records. Visible and near-ultraviolet 
data were
obtained from Johnson UBVRI measurements \citep{Johnson1966, Mendoza1967, 
Morel1978, Ducati2002} and from
\citet{Slesnick2002}. In addition we used observations in the Geneva 
\citep{Rufener1999}, Str\"omgren {\it uvby$\beta$} \citep{Marco2001} and 
Vilnius UPXYZVS systems \citep{Straizys1995}. The near-infrared
data (JHKs) was obtained from \citet{Morel1978} and \citet{Ducati2002}, as well 
as 2MASS \citep{Currie2010, 
Skrutskie2006}. Because the 2MASS data is saturated for both stars, the 
corresponding JHK fluxes are estimated using the less accurate profile method. 
Mid-infrared data was acquired by ISO-SWS \citep{Sloan2003}, AKARI/IRC
\citep{Ishihara2010}, WISE \citep{Wright2010,Cutri2012}, IRAS/LRS
\citep{Neugebauer1984}, and the Midcourse Space Experiment Point Source
Catalog \citep{Egan2003}. Far-infrared data came from IRAS
\citep{Neugebauer1984} and AKARI/FIS \citep{Ishihara2010} observations. 

The SEDs were de-reddened by adopting the extinction parameters found for the
Double Cluster in \citet{Slesnick2002}, i.e. a distance modulus of $11.85 \pm
0.05$ and color excesses of $E(B-V) = 0.53 \pm 0.02$ for RS~Per and $E(B-V) =
0.56 \pm 0.02$ for T~Per. Based on the analysis of \citet{McCall2004} and the
recommendations of \citet{Massey2005}, we attempted to de-redden the data
using two different empirical laws for the reddening curves: first using the
reddening curve from \citet{Cardelli1989} updated in the near-uv with
coefficients from \citet{ODonnell1994}, and with a total-to-selective ratio of
absorption $R_V= 4.15$ ; and second using the curve derived by
\citet{Fitzpatrick1999} with $R_V = 3.8$. In both cases we extended the
de-reddening to the longest wavelengths, using equations from
\citet{Chiar2006} for $\lambda > 5\mu$m. Figure~\ref{fig:seds} presents the
final de-reddened SEDs using the Cardelli de-reddenning.

Both SEDs show significant amount of circumstellar emission in the mid-IR and
far-IR, though is this much more pronounced for RS~Per. The infrared excess of
RS~Per around $7.6 \mu$m is attributed to PAH emission \citep{Verhoelst2009},
and a peak at 9.7 $\mu$m indicate strong silicate emission
\citep{Speck2000}. Both constitute evidence of ongoing dust
production. Moreover its far-infrared excess at 60$\mu$m is characteristic of
extended circumstellar emission \citep{Stencel1988, Stencel1989} and indicates
significant mass-loss through a circumstellar outflow whose typical size can
be estimated to about 4 arcminutes \citep{Stencel1989}. In contrast, T~Per
does not appear as active, but still displays broad dust emission between 9-13
$\mu$m range.

To determine the effective surface temperature, integration of the
spectrophotometric data has to be restricted to photospheric emission. To this
purpose we fitted SATLAS models only to the ultraviolet, visible and
near-infrared wavelengths where the stellar photosphere clearly dominates the
emission (i.e. $< 3 \mu$m). The SATLAS models were using ODF sampling with
improved H2O lines and the following parameters: sub-solar metallicity ${[\rm
    Fe/Z]} = -0.5$, based on the assumed typical metallicity of the double
cluster ${[\rm Fe/Z]} = -0.35$ from \citet{Gonzalez2000}; a medium
micro-turbulence level $\chi_t = 5$~km/s ; surface gravities $\log g = -0.5$
to $0.5$, and total mass $7-25 M_{\odot}$, typical of M~supergiants; effective
temperatures range $T = 3300-4000$~K, based on previous estimates
\citep{Gonzalez2000, Levesque2005, Verhoelst2009}. In order to check the 
independence of our results from the specificities of SATLAS, we also fitted 
spherical MARCS models using the same parameters, and we obtained identical fit 
results. We
found that the SED of T~Per was fitted well by models with temperatures in the
3700K--3800K range and $\log g\simeq 0$; and for RS~Per, temperature of
3500K--3600K and $\log g \simeq -0.5$.

To derive the bolometric fluxes, the photosphere SEDs was integrated with a
Gaussian quadrature algorithm (in logarithm space). The SATLAS model was used
instead of the actual SED only for wavelengths affected by circumstellar
emission (i.e. $> 3 \mu$m). Our estimates of bolometric fluxes and the derived
absolute bolometric magnitude) are presented in Table~\ref{tab:sed_fits}. The
1-$\sigma$ errors are mostly dominated by the uncertainties arising from the
de-reddening parameters, and from the probable inclusion of circumstellar
emission. We note that our estimate of RS~Per's absolute bolometric magnitude,
$\text{Mbol}=-7.47 \pm 0.12$, falls in the middle range of literature values:
$-7.21$ in \citet{Slesnick2002}, $-7.48$ in \citet{Gonzalez2000}, $-7.74$ in
\citet{Verhoelst2009}, and $-8.15$ in \citet{Levesque2005}.

\begin{figure}
\centering
\includegraphics[width=\linewidth]{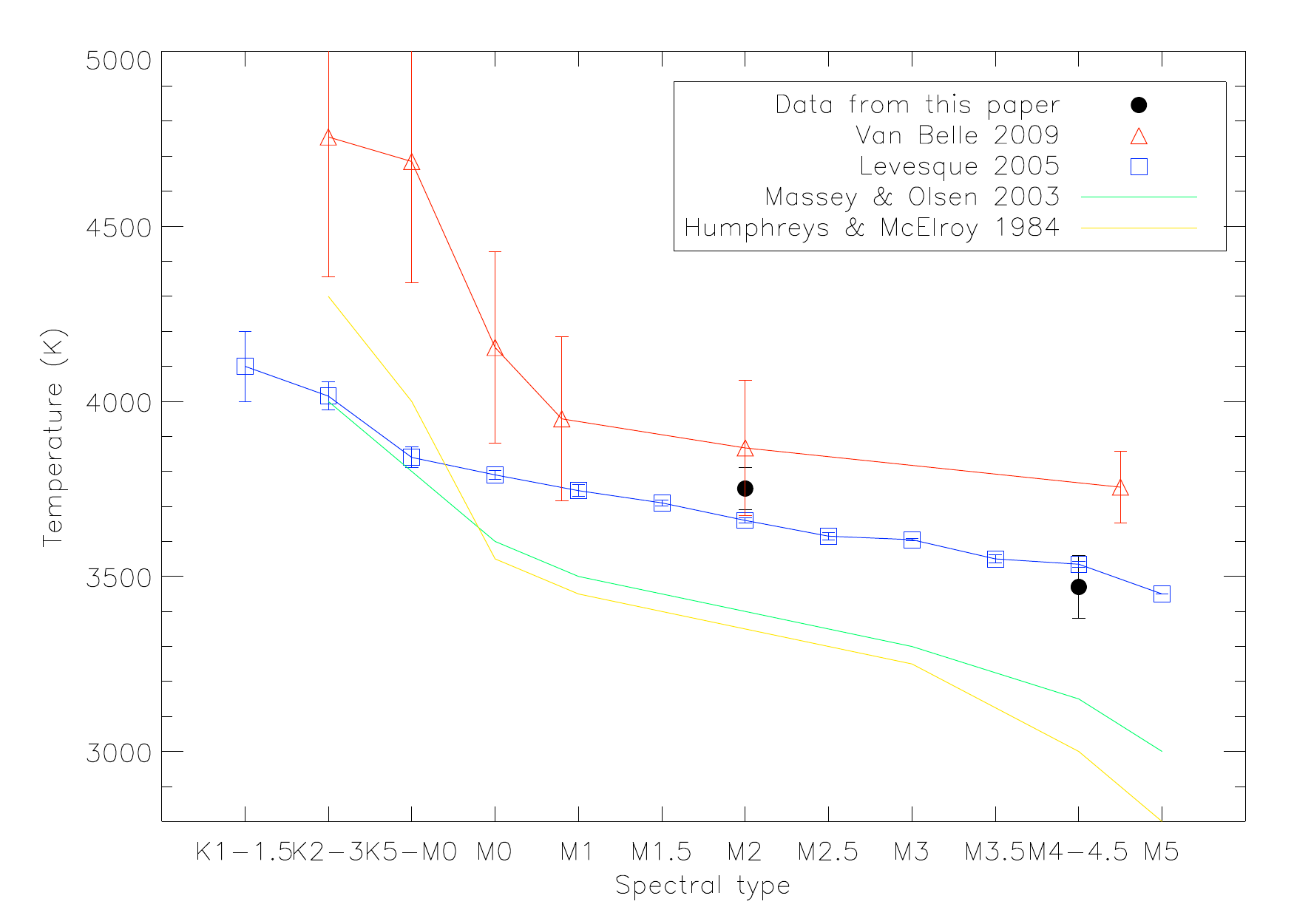}
\caption{Effective temperatures for T~Per and RS~Per, compared to four
temperatures scales from the literature. Our results confirm the hotter scale 
of \citet{Levesque2005} and \citet{vanBelle2009}.}\label{fig:tempscales}
\end{figure}

Using Equation~\ref{eq:temperature}, we derive the effective temperatures $T =
3685 \pm 30~K$ for T~Per and $T= 3470 \pm 90~K$ for RS~Per. Our error bars
take into account both the spectrophotometric and interferometric errors. The
dominant uncertainties lie in interstellar reddening: though relatively
well-characterized, the extinction to the Double Cluster is considerable due
to its distance, and the choice of de-reddening law significantly affect the
bolometric magnitude. These temperatures are consistent with previous
literature estimates \citep{Gonzalez2000, Slesnick2002,
  Verhoelst2009}. Figure~\ref{fig:tempscales} presents our results in relation
with four RSG temperature scales from the literature. Both scales from
\citet{Humphreys1984} and \citet{Massey2003} were derived by averaging
previous observations from public catalogs, but they most likely suffered from
de-reddening issues, underestimating the RSG
temperatures. \citet{Levesque2005} used spherical-geometry MARCS models (with 
the
then-new opacity sampling method, later published in \citet{Gustafsson2008})
and improved de-reddening of the sources, reconciling the observations with
both predicted temperatures and evolutionary tracks. Finally,
\citet{vanBelle2009} estimated the temperature by fitting the SED with stellar
templates derived from \citet{Pickles1998} in place of synthetic models, and
independently estimated stellar diameters using the interferometer PTI
(Palomar Tested Interferometer). Note that the spatial resolution of our CHARA
observations is at least twice that of PTI, therefore it should provide more
reliable diameter estimates.

Our results confirm the hotter temperature scales of RSGs, falling in-between
the results of \citet{Levesque2005} and \citet{vanBelle2009} for T~Per, and 
slightly under the Levesque's curve for RS~Per. Hence, and taking into account 
the limits of our analysis (noisy 2007 data compared to current CHARA/MIRC 
data), we are reasonably confident in the quality of our temperature estimates. 

\begin{deluxetable}{ccc}
\centering
\tabletypesize{\scriptsize}
\tablecaption{Estimated physical parameters of T~Per and RS~Per.}
\tablewidth{0pt}
\tablehead{ & \colhead{T~Per} & \colhead{RS~Per} }
\startdata
$R_{\text{ross}}$ ($R_{\odot}$) & $ 510 \pm 20$ & $ 770 \pm 30 $ \\
$M_{\rm bol}$ & $-6.90 \pm 0.07$ & $-7.47\pm 0.12$ \\
$T_{\rm eff} (K)$ & $3750 \pm 60$ & $3470 \pm 90$ \\
$M_{\star} (M_{\odot})$ & 9-12 & 12-15 \\
$\log L/L_{\odot}$ & $4.66 \pm 0.04$ & $4.89 \pm 0.05$ \\ 
$\log g$ (cgs) & $0.06 \pm 0.05$ & $-0.2 \pm 0.05$ \\
\enddata 
\label{tab:sed_fits}
\end{deluxetable}

\subsection{Linear radii, luminosity, mass, and surface gravity}

Assuming a distance of $d = 2345 \pm 55$ pc \citep{Slesnick2002}, our
estimates of the linear radii are $R_{\text{ross}}=510 \pm 20 R_{\odot}$ for 
T~Per and $R_{\text{ross}}=770 \pm 30~R_{\odot}$ for RS~Per. This corresponds 
to luminosities $\log L/L_{\odot} = 4.66 \pm 0.04$ for T~Per and $\log
L/L_{\odot} = 4.90 \pm 0.05$ for RS~Per, comparable to that of $\alpha$~Ori.
To get estimates of the stellar masses, we used the most recent evolutionary
tracks from \citet{Ekstrom2012} and the new Geneva tracks from
\citet{Neugent2012}, that both demonstrated their (relative) reliability on
RSGs. The range of possible masses appears to be $M= 9-12 M_{\odot}$ for
T~Per, and $M= 12-15 M_{\odot}$ for RS~Per, which translates into $\log g =
0.06 \pm 0.05$ for T~Per and $\log g = - 0.2 \pm 0.05$ for RS~Per. These
results support the assumptions made during our selection of SATLAS models in
sections~\ref{sec:ldd} and~\ref{sec:temperatures}.

\section{Image reconstruction} \label{sec:image_reconstruction}

\subsection{Regularized maximum likelihood}

As four telescope data is secured on both objects, there exists enough phase
information in our data sets to attempt ``model-independent
image reconstruction''. Here the prefix ``model-independent'' signifies that
the image reconstruction process will not rely on a specific astrophysical
model. The target image $\V{i}$ is modeled as an array of pixel
fluxes~$\widehat{\V{i}}=\{i_0, \ldots i_{n-1}\}$. 

As the data is assumed to be normally distributed, to each image we can
associate a $\chi^2(\V{i})$ metric that measure the distance between the
observed data (power spectra and bispectra) and the same quantities derived
from the current image. Maximizing the likelihood of the image by minimizing
its $\chi^2$ unfortunately does not lead to reasonable images. The reason is
that image reconstruction belongs to the class of ``ill-posed'' inverse
problems: the number of pixels to reconstruct is typically a few thousand,
while we only have a few hundreds interferometric data points. Under these
conditions, maximum likelihood leads to an overfitting the data. It is thus
essential to ``regularize'' the solution by introducing reasonable but
noncommittal prior expectations about the image. This is usually done through
regularization functions that control the flux distribution within the image.
In addition to preventing over-fitting, good regularizers fulfill other
roles. As underlined during model-fitting, the $\chi^2$ is heavily
multimodal. Most classic regularization function $R(\V{i})$ effectively allow
to discriminate between these local minima and thus ease minimization of
Eq.~\ref{eq:reconst}. In effect, regularizers help extrapolate the missing
information from the phase lost to the atmosphere and the gaps in the data
coverage of the ({\it u,v}) plane. A competent choice of regularizers ensures
that high frequencies are extrapolated well, and image reconstruction has
demonstrated it achieves super-resolution \citep{Renard2011}, i.e. that the
effective resolution of the reconstructed images is typically about three to
four times greater than the interferometer resolution.

This regularized maximum likelihood approach constitutes the current framework
for image reconstruction in optical interferometry \citep{Baron2010,
  Thiebaut2010}. Formally, the target image minimizes the sum of the
$\chi^2(\V{i})$ metric and of $K$~regularizers $R_k(\V{i})$:
\begin{equation}
\widehat{\V{i}} = \underset{i \in \mathbb{R}^n}{\operatorname{argmin}} \left \{ 
\chi^2 (\V{i}) + \sum_{k=1}^{K} \mu_k R_k(\V{i}) \right \}\label{eq:reconst} ,
\end{equation}
under the constraints of image positivity ($\forall n, i_n \ge 0$) and of
normalization of the image to unity ($\sum_n i_n = 1$). The
factors~$\mu_k$ in Eq.~\ref{eq:reconst} control the relative weight of the
$\chi^2$ and regularization terms.

\subsection{Reconstructing spotted stars with current 
software}\label{sec:oldreg}

Reconstructing spotted stars is currently difficult with available
software. To date, the only published model-independent interferometric
reconstructions of stellar spots are that of the large convection cells of
$\alpha$~Ori \citep{Young2000, Haubois2009, Chiavassa2010a} and of VX Sgr
\citep{Chiavassa2010b}. Resolving spots entails that the stellar disc is
proportionally much larger, which implies very low visibility amplitudes, and
consequently bad signal-to-noise. Moreover, the conventional convex
approximations of the $\chi^2$ expression should then be ruled out. And as
exemplified by the difficult reconstruction of VX Sgr, the minimization of the
non-convex $\chi^2$ is very prone to appearance of artifacts when using
conventional tools such as BSMEM~\citep{Baron2010} or
MIRA~\citep{Thiebaut2010}.
 
To solve this issue we suggest the use of non-convex reconstruction codes,
such as those based on Markov Chain Monte Carlo (MCMC) approach. In this
paper, the software SQUEEZE \citep{Baron2010} was used to obtain the
reconstructions presented in this paper. SQUEEZE uses parallel tempering to
tentatively find the global minimum of the criterion in Eq.~\ref{eq:reconst},
and therefore is well-adapted to non-convex problems. SQUEEZE is
multi-threaded, with each thread conducting minimization by simulated
annealing at a different temperature and starting with a different random
seed. Compared to its predecessor MACIM \citep{Ireland2006}, it is less
sensitive to the initial condition of the Markov Chains (i.e. the starting
image). Thus, the quality of its reconstructions mostly depends on the choice
of regularization.

\begin{figure}
\centering
\includegraphics[width=0.5\linewidth]{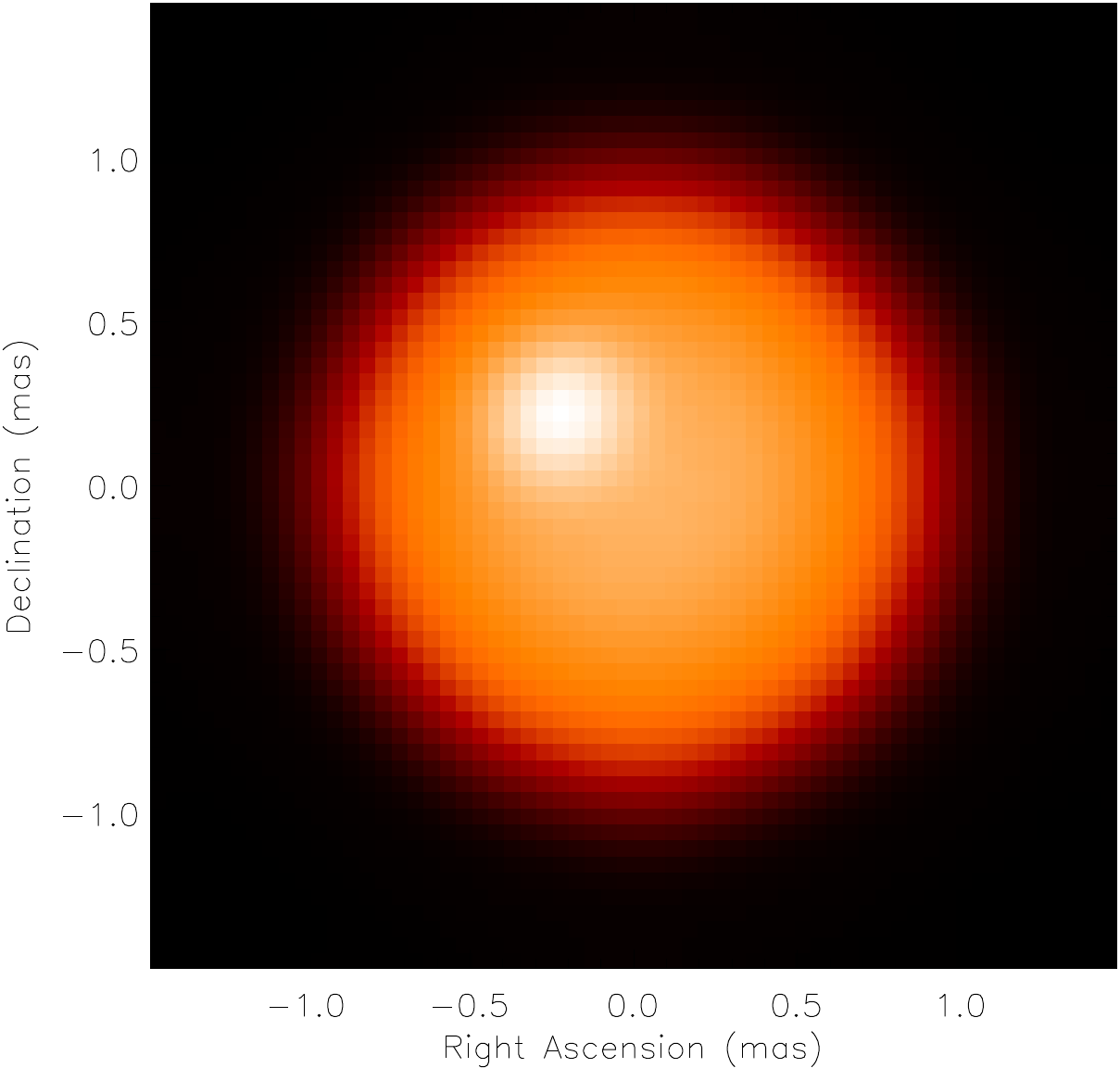}%
\includegraphics[width=0.5\linewidth]{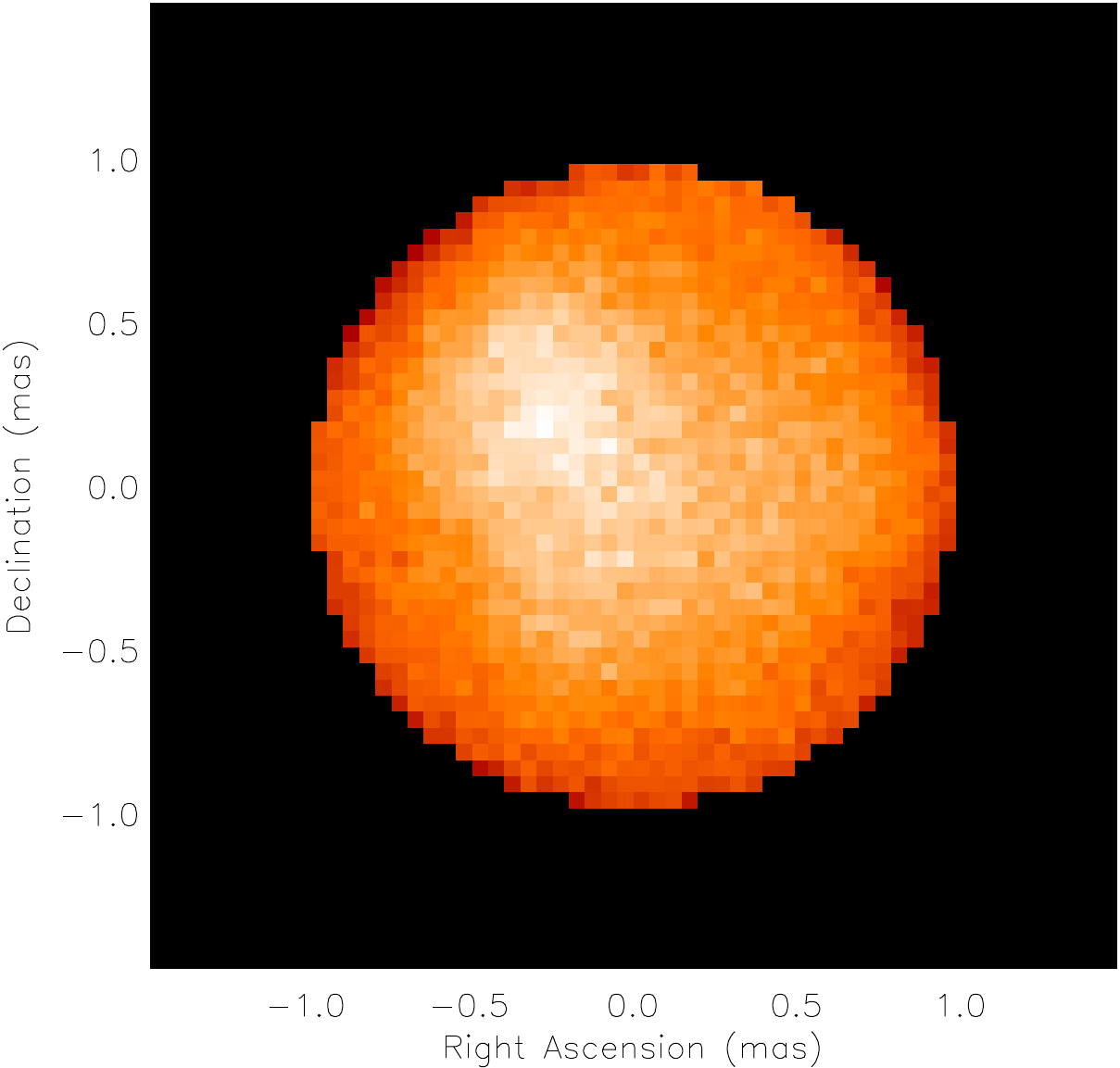}
\includegraphics[width=0.5\linewidth]{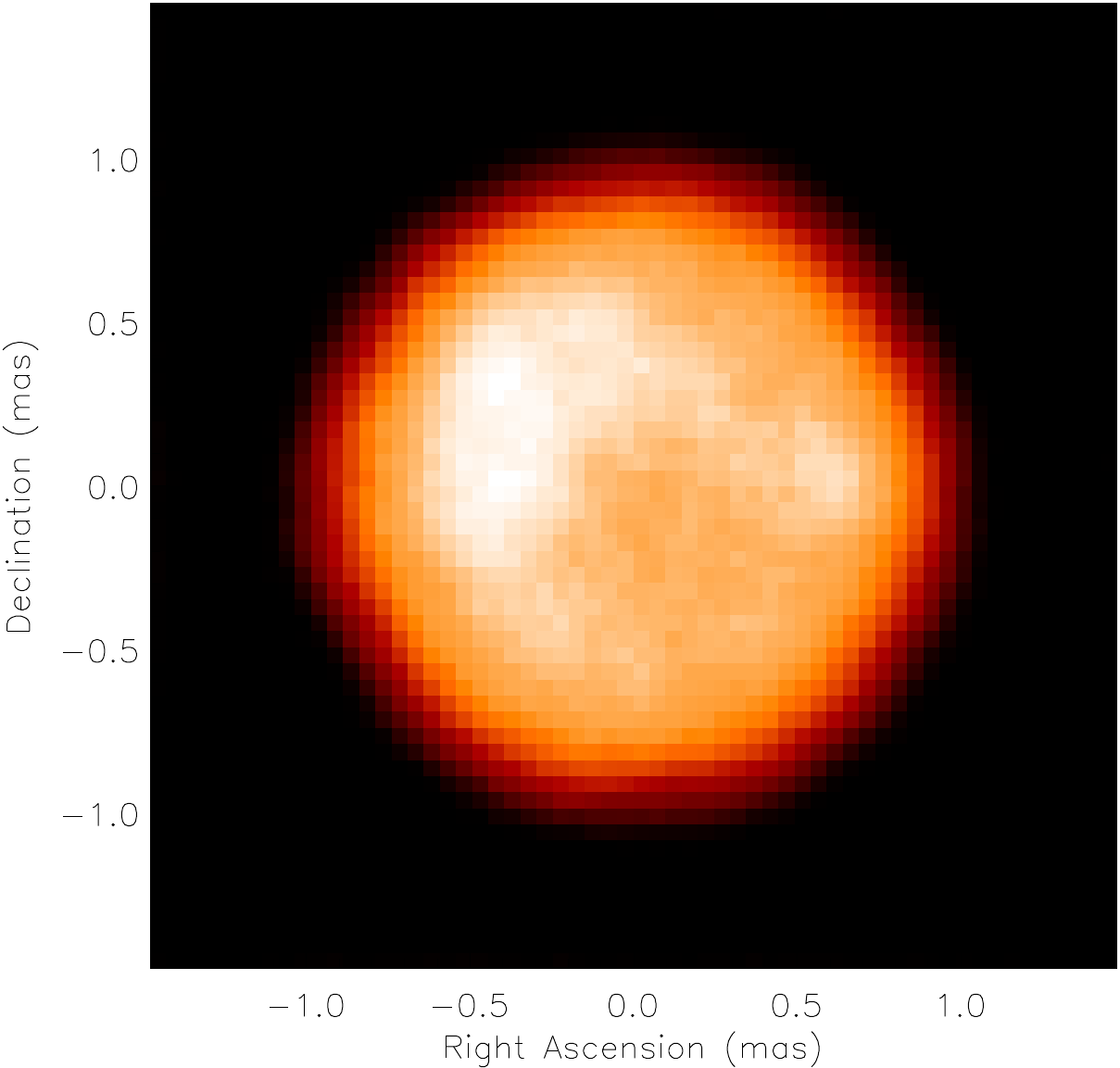}%
\includegraphics[width=0.5\linewidth]{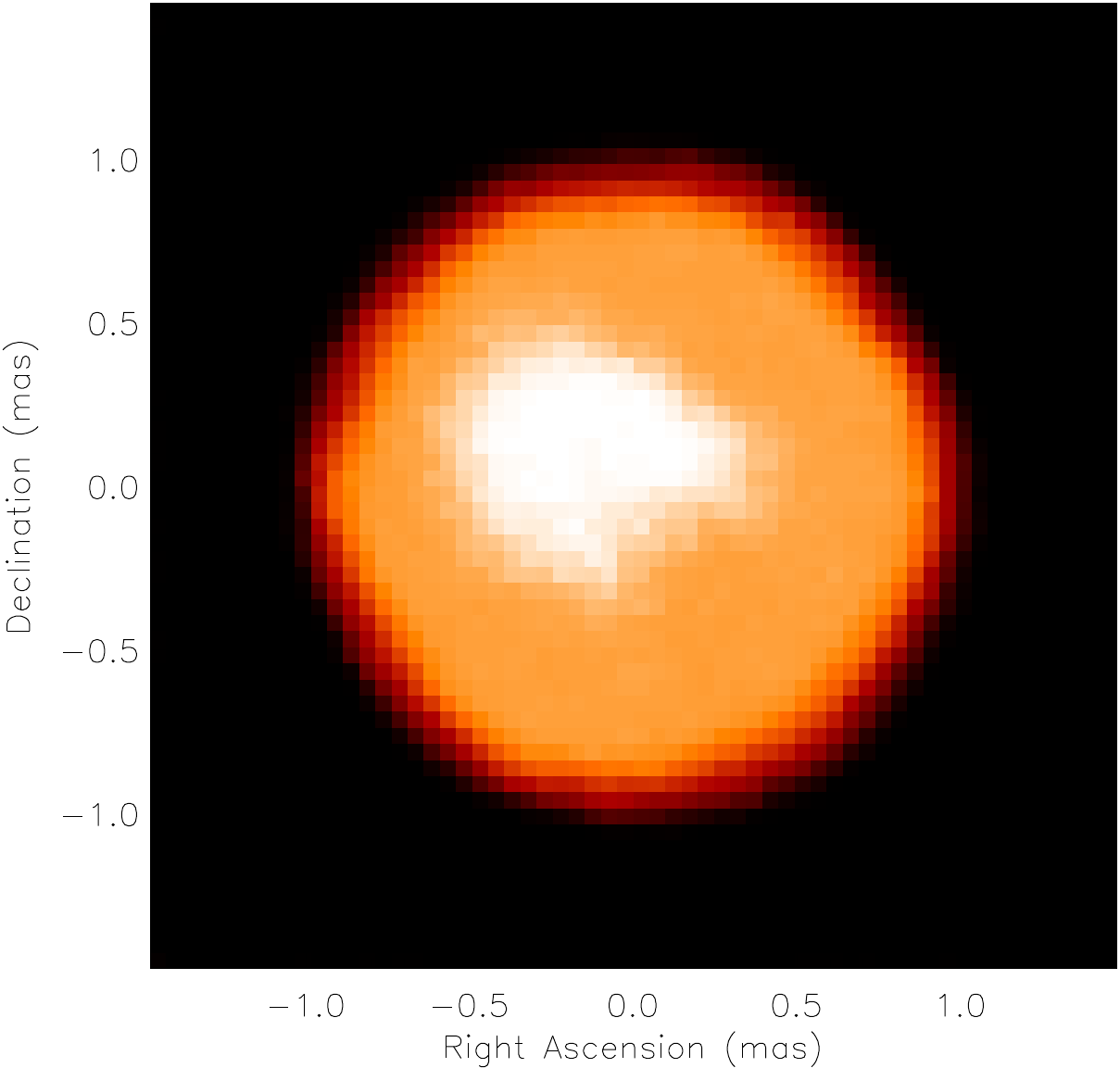}%
\caption{Reconstructions of a synthetic spotted star with the same same {uv} 
coverage and signal-to-noise as the T~Per data. Top left: the original image 
convolved to the expected
  effective resolution (using a super-resolution factor of~3); Top right:
  reconstruction regularized by maximum entropy and a prior constraining the
  flux to stay within the stellar diameter; Bottom left: reconstruction
  regularized by total variation ; Bottom right: reconstruction regularized by
  the spot regularizer presented in
  Section~\ref{sec:newreg}. }\label{fig:reconst_compare}
\end{figure}

To select the best regularizer, we generated a synthetic test dataset
simulating the observation of a spotted star using the OIFITS-SIM tools
\citep{oifitssim}. The original image used to create the data was chosen as
the T~Per "bright spot" model from Figure~\ref{fig:tper_spots}, and we use the
same ({\it u,v}) coverage and signal-to-noise as the actual T~Per data.  We
then reconstructed the stellar surface using the two most successful
regularizers as benchmarked by \citet{Renard2011}: maximum entropy and total
variation~\citep{Rudin1992}. Maximum entropy was implemented using the
multiplicity expression from \citet{Sutton2006}, which is well adapted to our
MCMC implementation:
\begin{equation}
R_{\Gamma}(\V{i}) = \sum_{n} \log \Gamma(i_n +1).
\end{equation}
where $i_n$ is the flux in pixel $n$. Total variation (hereafter, TV) is 
defined as the $\ell_1$ norm of the spatial gradient $\V{g}$:
\begin{equation}
R_{\text{TV}}(\V{i}) = \ell_1(\V{g}) = \sum_{n} |g_{n}| .
\end{equation}
Several practical expressions are available to discretize $\V{g}$ on the image 
grid. In the context of this paper, we implemented the classic isotropic 
formulation of $\V{g}$, i.e. for each pixel coordinate $(n, m)$ in the 
two-dimensional image~$\V{i}$, the local gradient was given by:
\begin{equation}
g_{n,m}(\V{i}) = \sqrt{|i_{n+1, m}-i_{n,m}|^2 + |i_{n, m+1}-i_{n,m}|^2}.
\end{equation} 
Figure~\ref{fig:reconst_compare} compares regularization obtained with both
these regularizers on our synthetic dataset (the full reconstruction procedure
is detailed in Section~\ref{sec:reconst_proc}). Our results demonstrate that
the maximum entropy image suffers from several flaws: the stellar background
is excessively non-uniform, and the precise location of spot is lost. The
total variation reconstruction is definitively superior on both
aspects. Moreover, and unlike maximum entropy, total variation does not
require an additional prior to constrain the flux to stay within a given
diameter. The good performance of total variation are in line with the
empirical results of \citet{Renard2011} but also theoretical
predictions. Total variation is indeed a direct application of the Compressed
Sensing theory, a recent mathematical framework that supersedes the
conventional Shannon sampling theorem when applied to sparse images, i.e.,
images that may be described with small number of non-zero coefficients in
some give basis. Here, on first order, our model spotted star consist of a
(mostly) uniform disc with compact spots or cells. The spatial gradient of the
image is sparse, with only the perimeters of the stellar disc and the spots as
non-zero components. Total variation enforces the sparsity of the spatial
gradient so that the reconstruction is piecewise constant with sharp
transitions, though this is not apparent on Figure~\ref{fig:reconst_compare}
as these images are actually Markov Chain averages as explained further in
Section~\ref{sec:reconst_proc}.

\subsection{A novel regularizer for spotted stars}\label{sec:newreg}

To derive a novel regularizer for use on spotted stars, we built upon the idea
of spatial gradient sparsity, adding two noncommittal requirements. The first
requirement is that for a given flux to distribute into possible spots, the
regularizer should prefer a single spot to two spots, as per Occam's razor
prescription. The second requirement is that for a given flux to attribute to
a spot, the size of the spot should be determined solely by the data. This
implies that the regularizer value should be independent of the spot size.

It is straightforward to demonstrate that TV violates this second
requirement. Let us consider an idealized stellar disc, and two cases: either
two small spots of diameter $D$, or a single large spot of diameter $2D$. Let
us assume the brightness distribution of the disc and the spots as uniform, so
that the spatial gradient is null everywhere except on the perimeters of these
components. The actual contribution of the spots to the spatial gradient is
then proportional to the spot perimeters -- equal to $2 \pi D$ in both cases
-- multiplied by the flux density for each case. Assuming that a combined flux
$F$ is emitted by the spots, the spot flux density is then $F / ( \pi D^2)$
for the single spot case, and $2 F / ( \pi D^2)$ for the two spot
case. Consequently the total variation is $\operatorname{TV}(\text{one spot})
= F/D$ for a single spot and $\operatorname{TV}(\text{2 spots}) = 2F/D$ for
two spots. While this implies that TV does favor a single spot, this also
demonstrates that the regularization depends on the size of the spot, and
therefore it may bias a reconstruction toward larger spots.

In contrast the regularizer $R_{\text{spot}}$ defined by:
\begin{equation}
R_{\text{spot}}(\V{i}) = \ell_{\frac{1}{2}}(\V{g}) = \left(\sum_{n} 
\sqrt{|g_{n}|} \right)^2 .
\end{equation}
meets both our requirements for an ideal regularizer, with 
$R_{\text{spot}}(\text{one spot}) = 4 \pi F$ and $R_{\text{spot}}(\text{two 
spots}) = 8 \pi F$. 
Figure~\ref{fig:reconst_compare} confirms our analysis, and our spot 
regularizer demonstrates a significant improvement over total variation. 

\subsection{Reconstruction procedure and results}\label{sec:reconst_proc}

\begin{figure}
\centering
\includegraphics[width=0.5\linewidth]{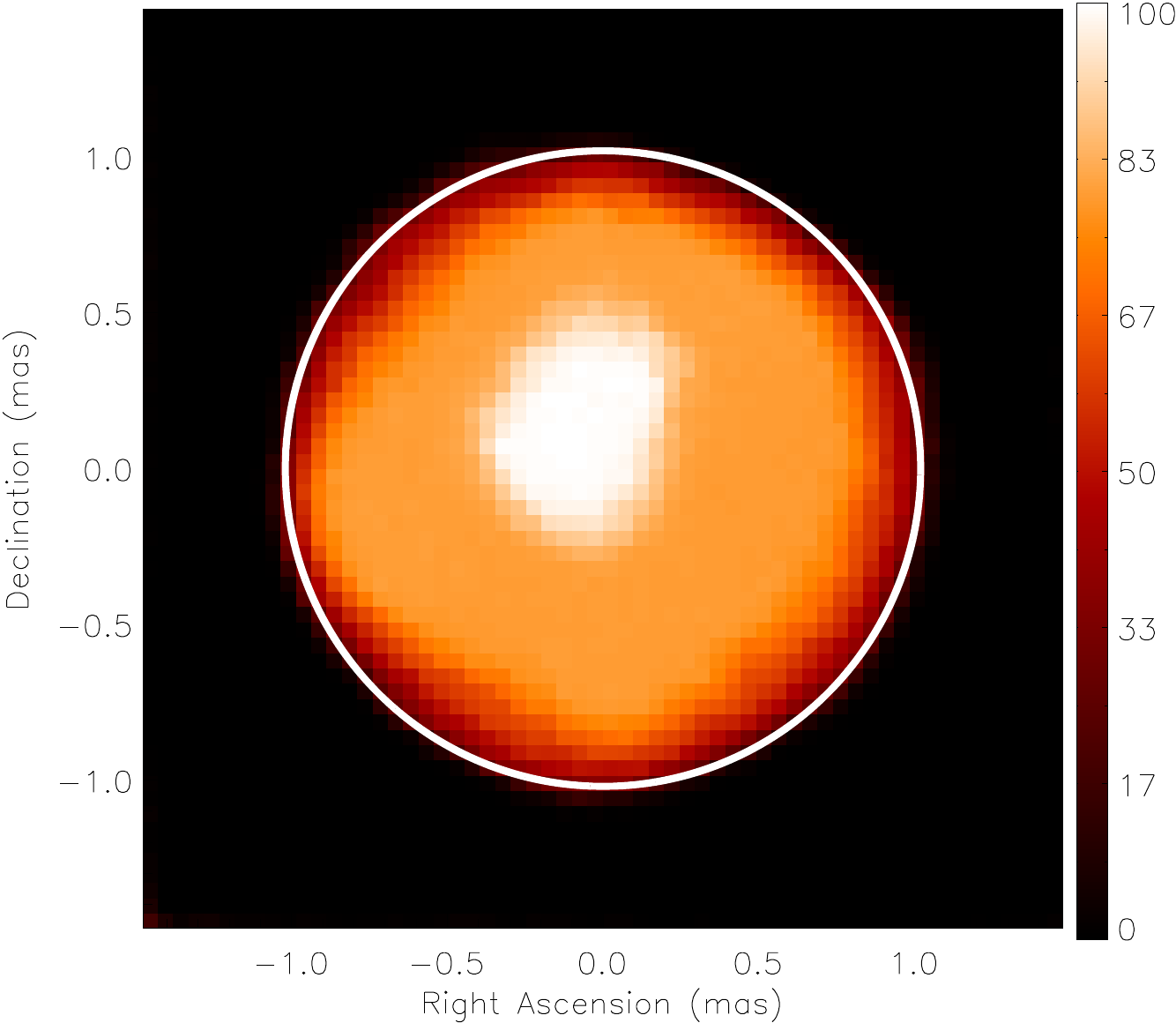}%
\includegraphics[width=0.5\linewidth]{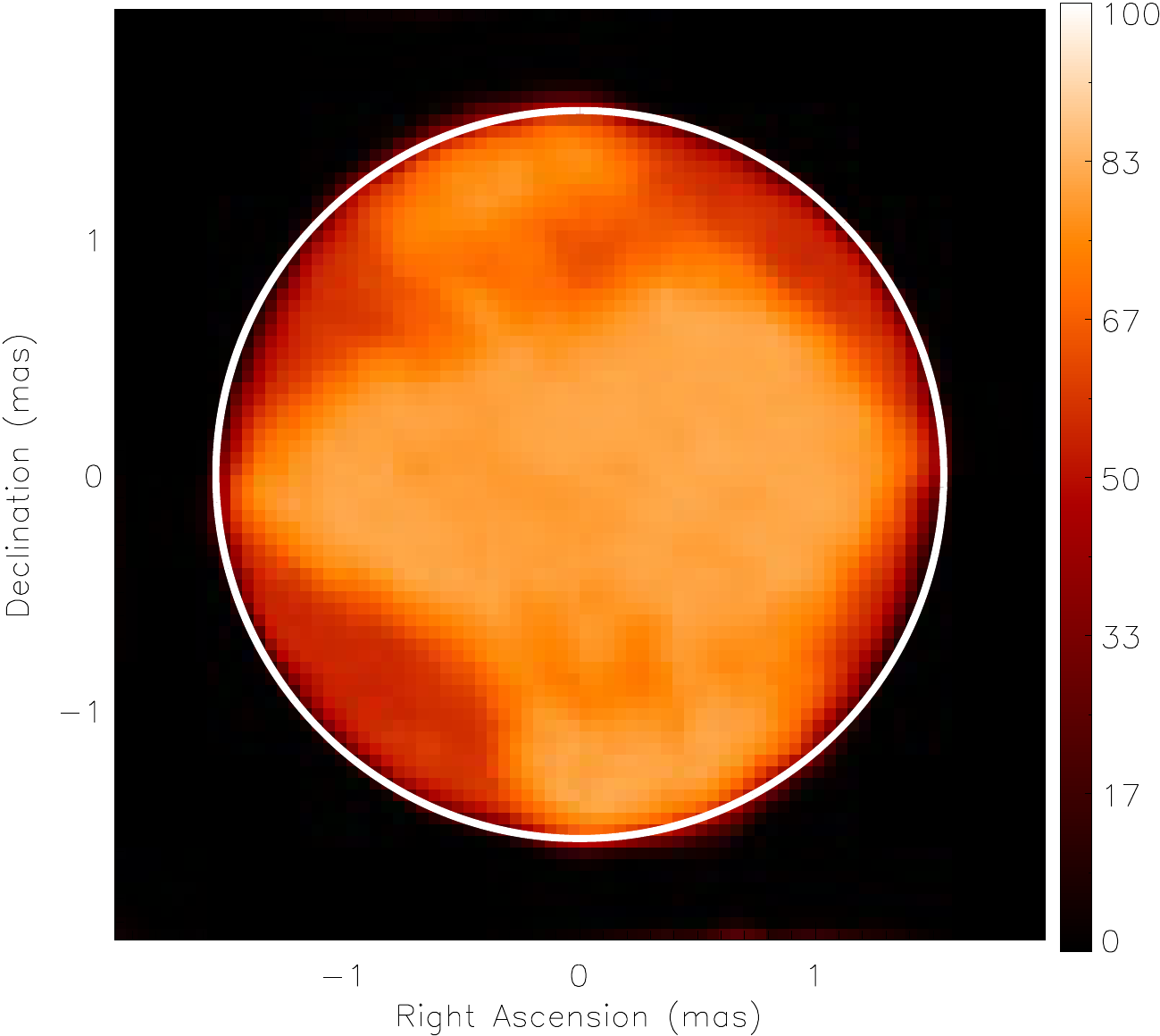}%
\caption{Reconstructed images of T~Per (left) and RS~Per (right) with the
  SQUEEZE-MCMC engine and the ''spot regularizer'' presented in
  section~\ref{sec:newreg}. The angular diameters estimated by model-fitting
  are indicated by white circles.}\label{fig:images}
\end{figure}

The instrumental resolution is given by the largest CHARA baseline in our data
sets (S2-W1 or E2-W1, $\simeq 250$ meters), corresponding to~$1.3$ mas in H
band. Taking into account a super-resolution factor of four, the effective
resolution of the reconstructed images should be about $0.3$~mas. In order to
avoid excessive pixellation of the images, the actual resolution of the
reconstruction was set to $0.1$~mas. We ran five batches of multi-threaded
SQUEEZE with $16$~threads each, corresponding to a total of $80$~independent
Markov chains that were averaged to reconstruct the final images. The number
of pixel elements in each chain was set to~$5000$, with a length of~$500$
iterations. In addition to the spot regularizer derived in the previous
section, we made use of the fitting results from section~\ref{sec:ldd} to
constrain the reconstruction. The fitted limb-darkening discs were used to
initialize the chains to sensible starting points.The factors~$\mu_k$ were
chosen so that the actual reduced $\chi^2$ is roughly unity for the
reconstructed image.  The final reconstructions are presented on
Figure~\ref{fig:images}. For T~Per, the spot location in North West quadrant
agrees with the conclusions of Bayesian model selection. For RS~Per, there is
indeed a darker area in the South West. However the correct interpretation is
unclear: this may be a dark spot, or most of the surface could be understood
as a temporary hot convection cell. Without data outside the 2007 July/August
period, we cannot conclude from this single RS~Per image. To exclude the 
possibility that the surface
features on Figure~\ref{fig:images} are due to ({\it u,v}) coverage or to
noisy data, we ran an ''artifact test" on both targets. We generated synthetic
observations of the limb-darkening discs derived from model fitting, with
exactly the same ({\it u, v}) coverage and signal-to-noise as the real data
sets. We then reconstructed the images using the same procedure outlined
above, and we found that the reconstructions did not display any significant
surface features.

\section{Discussion and Conclusion}

We have analyzed the CHARA/MIRC data taken in H~band on two red supergiants
T~Per and RS~Per and presented evidence for the existence of spots on their
surfaces. For this we developed a set of tools dedicated to the analysis of
spotted stars. As the utility of simple model-fitting procedures is limited
for spotted stars, we demonstrated that Bayesian model selection is capable of
assessing the relative probabilities of various models, the Bayesian evidence
constituting a more reliable metric than the reduced $\chi^2$. Our results on
T~Per confirm that hot spots can indeed be observed in $H$-band. If considered 
together with similar results on $\alpha$~Ori by \citep{Haubois2009} and on 
VX~Sgr by \citep{Chiavassa2010a}, it seems we should expect a significant 
proportion of RSGs to have bright spots, as is thought to be the case for AGB 
stars \citep{Ragland2006}. Considering the typical continuum opacity curves in 
such cool atmospheres \citep{Woodruff2009}, the continuum opacity should be 
close to 
minimum in H~band, where our MIRC observations took place. These spots must be 
generated very 
close to the photosphere, and therefore it seems currently doubtful that their 
enhanced contrast may be explained by opacity effects. 
As the correct approach to model these objects is not really to model 
spots, but to interpret the surface in terms of convective cells using 3D 
models \citep{Chiavassa2010b}, inhomogeneous granulation temperatures may 
explain the 
spots. Our detection of a ''dark spot`` on RS~Per probably corresponds to a 
cooler 
granulation, imaged with the reduced dynamic contrast typical of current 
interferometry.

We also found that classic regularizers are hardly adequate to reconstruct
model-independent images of spotted surfaces. Hence we derived a novel
regularizer tailored for this task, based on simple Compressed Sensing and
Occam's razor principles. Our reconstructions of T~Per and RS~Per were found
to essentially agree with the Bayesian spot selection.

It should be underlined that the data quality from MIRC circa 2007 was a major
limiting factor in the present analysis. Fortunately since 2007, the MIRC
combiner underwent a series of hardware upgrades which drastically improved
its performance. MIRC-6T can now simultaneously combine all six CHARA
telescopes with thrice higher signal-to-noise and much lower systematic
errors. A survey of several RSGs over longer periods of time with MIRC-6T
would allow to conclude to whether spots on M~supergiants are ubiquitous, and
in particular if a relationship can be found between circumstellar
activity/infrared excess and the complexity of observed surface features. Our
future work will thus focus on the analysis of new RSG data collected with
MIRC-6T, with a systematic application of Bayesian model selection to
3D~hydrodynamical models, and the development of better reconstruction
algorithms for RSGs.

\acknowledgments

The authors would like to acknowledge funding from the NSF through award
AST-0807577 to the University of Michigan, as well as funding by the
Australian Research Council and the ``Lend\"ulet Fiatal Kutat\'oi Program'' of
the Hungarian Academy of Sciences for the initial work on these objects.

This work is based on observations collected at the Center for High Angular
Resolution Astronomy (CHARA) operated by the Georgia State University at
Mt. Wilson, California. This research also relied on observations with AKARI,
a JAXA project with the participation of ESA, and made use of data products
from the Wide-field Infrared Survey Explorer (a joint project of the
University of California, Los Angeles, and the Jet Propulsion
Laboratory/California Institute of Technology, funded by the National
Aeronautics and Space Administration), and from the Two Micron All Sky Survey
(a joint project of the University of Massachusetts and the Infrared
Processing and Analysis Center/California Institute of Technology, funded by
the National Aeronautics and Space Administration and the National Science
Foundation).


\begin{thebibliography}{103}
\expandafter\ifx\csname natexlab\endcsname\relax\def\natexlab#1{#1}\fi

\bibitem[{{Airapetian} {et~al.}(2010){Airapetian}, {Carpenter}, \&
  {Ofman}}]{Airapetian2010}
{Airapetian}, V., {Carpenter}, K.~G., \& {Ofman}, L. 2010, \apj, 723, 1210

\bibitem[{{Auri{\`e}re} {et~al.}(2010){Auri{\`e}re}, {Donati},
  {Konstantinova-Antova}, {Perrin}, {Petit}, \& {Roudier}}]{Auriere2010}
{Auri{\`e}re}, M., {Donati}, J.-F., {Konstantinova-Antova}, R., {Perrin}, G.,
  {Petit}, P., \& {Roudier}, T. 2010, \aap, 516, L2

\bibitem[{Baron {et~al.}(2010)Baron, Monnier, \& Kloppenborg}]{Baron2010}
Baron, F., Monnier, J., \& Kloppenborg, B. 2010, \procspie, 7734, 77342I

\bibitem[{{Buscher} {et~al.}(1990){Buscher}, {Baldwin}, {Warner}, \&
  {Haniff}}]{Buscher1990}
{Buscher}, D.~F., {Baldwin}, J.~E., {Warner}, P.~J., \& {Haniff}, C.~A. 1990,
  \mnras, 245, 7P

\bibitem[{{Cardelli} {et~al.}(1989){Cardelli}, {Clayton}, \&
  {Mathis}}]{Cardelli1989}
{Cardelli}, J.~A., {Clayton}, G.~C., \& {Mathis}, J.~S. 1989, \apj, 345, 245

\bibitem[{{Castelli} \& {Kurucz}(2003)}]{Castelli2004}
{Castelli}, F., \& {Kurucz}, R.~L. 2003, in IAU Symposium, Vol. 210, Modelling
  of Stellar Atmospheres, ed. N.~{Piskunov}, W.~W. {Weiss}, \& D.~F. {Gray},
  20P

\bibitem[{{Chiar} \& {Tielens}(2006)}]{Chiar2006}
{Chiar}, J.~E., \& {Tielens}, A.~G.~G.~M. 2006, \apj, 637, 774

\bibitem[{{Chiavassa} {et~al.}(2011){Chiavassa}, {Freytag}, {Masseron}, \&
  {Plez}}]{Chiavassa2011}
{Chiavassa}, A., {Freytag}, B., {Masseron}, T., \& {Plez}, B. 2011, \aap, 535,
  A22

\bibitem[{{Chiavassa} {et~al.}(2010{\natexlab{a}}){Chiavassa}, {Haubois},
  {Young}, {Plez}, {Josselin}, {Perrin}, \& {Freytag}}]{Chiavassa2010b}
{Chiavassa}, A., {Haubois}, X., {Young}, J.~S., {Plez}, B., {Josselin}, E.,
  {Perrin}, G., \& {Freytag}, B. 2010{\natexlab{a}}, \aap, 515, A12

\bibitem[{{Chiavassa} {et~al.}(2009){Chiavassa}, {Plez}, {Josselin}, \&
  {Freytag}}]{Chiavassa2009}
{Chiavassa}, A., {Plez}, B., {Josselin}, E., \& {Freytag}, B. 2009, \aap, 506,
  1351

\bibitem[{{Chiavassa} {et~al.}(2010{\natexlab{b}}){Chiavassa}, {Lacour},
  {Millour}, {Driebe}, {Wittkowski}, {Plez}, {Thi{\'e}baut}, {Josselin},
  {Freytag}, {Scholz}, \& {Haubois}}]{Chiavassa2010a}
{Chiavassa}, A., {et~al.} 2010{\natexlab{b}}, \aap, 511, A51

\bibitem[{{Cuntz}(1997)}]{Cuntz1997}
{Cuntz}, M. 1997, \aap, 325, 709

\bibitem[{{Currie} {et~al.}(2010){Currie}, {Hernandez}, {Irwin}, {Kenyon},
  {Tokarz}, {Balog}, {Bragg}, {Berlind}, \& {Calkins}}]{Currie2010}
{Currie}, T., {et~al.} 2010, \apjs, 186, 191

\bibitem[{{Cutri} {et~al.}(2012){Cutri}, {Wright}, {Conrow}, {Bauer},
  {Benford}, {Brandenburg}, {Dailey}, {Eisenhardt}, {Evans}, {Fajardo-Acosta},
  {Fowler}, {Gelino}, {Grillmair}, {Harbut}, {Hoffman}, {Jarrett},
  {Kirkpatrick}, {Leisawitz}, {Liu}, {Mainzer}, {Marsh}, {Masci}, {McCallon},
  {Padgett}, {Ressler}, {Royer}, {Skrutskie}, {Stanford}, {Wyatt}, {Tholen},
  {Tsai}, {Wachter}, {Wheelock}, {Yan}, {Alles}, {Beck}, {Grav}, {Masiero},
  {McCollum}, {McGehee}, {Papin}, \& {Wittman}}]{Cutri2012}
{Cutri}, R.~M., {et~al.} 2012, {Explanatory Supplement to the WISE All-Sky Data
  Release Products}, Tech. rep.

\bibitem[{{Danchi} {et~al.}(1994){Danchi}, {Bester}, {Degiacomi}, {Greenhill},
  \& {Townes}}]{Danchi1994}
{Danchi}, W.~C., {Bester}, M., {Degiacomi}, C.~G., {Greenhill}, L.~J., \&
  {Townes}, C.~H. 1994, \aj, 107, 1469

\bibitem[{{Ducati}(2002)}]{Ducati2002}
{Ducati}, J.~R. 2002, VizieR Online Data Catalog, 2237, 0

\bibitem[{{Egan} {et~al.}(2003){Egan}, {Price}, {Kraemer}, {Mizuno}, {Carey},
  {Wright}, {Engelke}, {Cohen}, \& {Gugliotti}}]{Egan2003}
{Egan}, M.~P., {et~al.} 2003, VizieR Online Data Catalog, 5114, 0

\bibitem[{{Ekstr{\"o}m} {et~al.}(2012){Ekstr{\"o}m}, {Georgy}, {Eggenberger},
  {Meynet}, {Mowlavi}, {Wyttenbach}, {Granada}, {Decressin}, {Hirschi},
  {Frischknecht}, {Charbonnel}, \& {Maeder}}]{Ekstrom2012}
{Ekstr{\"o}m}, S., {et~al.} 2012, \aap, 537, A146

\bibitem[{{Feroz} \& {Hobson}(2008)}]{Feroz2008}
{Feroz}, F., \& {Hobson}, M.~P. 2008, \mnras, 384, 449

\bibitem[{{Feroz} {et~al.}(2009){Feroz}, {Hobson}, \& {Bridges}}]{Feroz2009}
{Feroz}, F., {Hobson}, M.~P., \& {Bridges}, M. 2009, \mnras, 398, 1601

\bibitem[{{Fitzpatrick}(1999)}]{Fitzpatrick1999}
{Fitzpatrick}, E.~L. 1999, \pasp, 111, 63

\bibitem[{Freytag {et~al.}(2002)Freytag, Steffen, \& Dorch}]{Freytag2002}
Freytag, B., Steffen, M., \& Dorch, B. 2002, Astronom. Nach., 323, 213

\bibitem[{Gonzalez \& Wallerstein(2000)}]{Gonzalez2000}
Gonzalez, G., \& Wallerstein, G. 2000, \apj, 119, 1839

\bibitem[{{Gray}(2008)}]{Gray2008}
{Gray}, D.~F. 2008, \aj, 135, 1450

\bibitem[{{Grunhut} {et~al.}(2010){Grunhut}, {Wade}, {Hanes}, \&
  {Alecian}}]{Grunhut2010}
{Grunhut}, J.~H., {Wade}, G.~A., {Hanes}, D.~A., \& {Alecian}, E. 2010, \mnras,
  408, 2290

\bibitem[{{Gustafsson} {et~al.}(1975){Gustafsson}, {Bell}, {Eriksson}, \&
  {Nordlund}}]{Gustafsson1975}
{Gustafsson}, B., {Bell}, R.~A., {Eriksson}, K., \& {Nordlund}, A. 1975, \aap,
  42, 407

\bibitem[{{Gustafsson} {et~al.}(2008){Gustafsson}, {Edvardsson}, {Eriksson},
  {J{\o}rgensen}, {Nordlund}, \& {Plez}}]{Gustafsson2008}
{Gustafsson}, B., {Edvardsson}, B., {Eriksson}, K., {J{\o}rgensen}, U.~G.,
  {Nordlund}, {\AA}., \& {Plez}, B. 2008, \aap, 486, 951

\bibitem[{{Haniff}(1991)}]{Haniff1991}
{Haniff}, C.~A. 1991, \josaa, 8, 134

\bibitem[{Haubois {et~al.}(2009)Haubois, Perrin, Lacour, Verhoelst, Meimon,
  Mugnier, Thi\'{e}baut, Berger, Ridgway, Monnier, Millan-Gabet, \&
  Traub}]{Haubois2009}
Haubois, X., {et~al.} 2009, \aap, 508, 923

\bibitem[{{Hauschildt} \& {Baron}(1999)}]{Hauschildt1999}
{Hauschildt}, P.~H., \& {Baron}, E. 1999, Journal of Computational and Applied
  Mathematics, 109, 41

\bibitem[{Hestroffer(1997)}]{Hestroffer1997}
Hestroffer, D. 1997, \aap, 327, 199

\bibitem[{Humphreys \& McElroy(1984)}]{Humphreys1984}
Humphreys, R., \& McElroy, D. 1984, \apj, 284, 565

\bibitem[{{Ireland} {et~al.}(2006){Ireland}, {Monnier}, \&
  {Thureau}}]{Ireland2006}
{Ireland}, M.~J., {Monnier}, J.~D., \& {Thureau}, N. 2006, in \procspie, Vol.
  6268

\bibitem[{{Ishihara} {et~al.}(2010){Ishihara}, {Onaka}, {Kataza}, {Salama},
  {Alfageme}, {Cassatella}, {Cox}, {Garcia-Lario}, {Stephenson}, {Cohen},
  {Fujishiro}, {Fujiwara}, {Hasegawa}, {Ita}, {Kim}, {Matsuhara}, {Murakami},
  {Muller}, {Nakagawa}, {Ohyama}, {Oyabu}, {Pyo}, {Sakon}, {Shibai}, {Takita},
  {Tanab}, {Uemizu}, {Ueno}, {Usui}, {Wada}, {Watarai}, {Yamamura}, \&
  {Yamauchi}}]{Ishihara2010}
{Ishihara}, D., {et~al.} 2010, VizieR Online Data Catalog, 2297, 0

\bibitem[{{Johnson} {et~al.}(1966){Johnson}, {Mitchell}, {Iriarte}, \&
  {Wisniewski}}]{Johnson1966}
{Johnson}, H.~L., {Mitchell}, R.~I., {Iriarte}, B., \& {Wisniewski}, W.~Z.
  1966, Communications of the Lunar and Planetary Laboratory, 4, 99

\bibitem[{{Josselin} \& {Plez}(2007)}]{Josselin2007}
{Josselin}, E., \& {Plez}, B. 2007, \aap, 469, 671

\bibitem[{Kass \& Raftery(1995)}]{Kass1995}
Kass, R.~E., \& Raftery, A.~E. 1995, Journal of the American Statistical
  Association, 90, 773

\bibitem[{{Kervella} \& {Fouqu{\'e}}(2008)}]{Kervella2008}
{Kervella}, P., \& {Fouqu{\'e}}, P. 2008, \aap, 491, 855

\bibitem[{{Kervella} {et~al.}(2011){Kervella}, {Perrin}, {Chiavassa},
  {Ridgway}, {Cami}, {Haubois}, \& {Verhoelst}}]{Kervella2011}
{Kervella}, P., {Perrin}, G., {Chiavassa}, A., {Ridgway}, S.~T., {Cami}, J.,
  {Haubois}, X., \& {Verhoelst}, T. 2011, \aap, 531, A117

\bibitem[{Kiss {et~al.}(2006)Kiss, Szab\'{o}, \& Bedding}]{Kiss2006}
Kiss, L.~L., Szab\'{o}, G.~M., \& Bedding, T.~R. 2006, \mnras, 372, 1721

\bibitem[{{Kloppenborg} \& {Baron}(2013)}]{oifitssim}
{Kloppenborg}, B., \& {Baron}, F. 2013, {OIFITS-SIM},
  \url{https://github.com/bkloppenborg/oifits-sim}, [Online]

\bibitem[{{Kurucz}(1992)}]{Kurucz1992}
{Kurucz}, R.~L. 1992, in IAU Symposium, Vol. 149, The Stellar Populations of
  Galaxies, ed. B.~{Barbuy} \& A.~{Renzini}, 225

\bibitem[{{Lester} \& {Neilson}(2008)}]{Lester2008}
{Lester}, J.~B., \& {Neilson}, H.~R. 2008, \aap, 491, 633

\bibitem[{{Levesque}(2010)}]{Levesque2010}
{Levesque}, E.~M. 2010, New Astronomy Reviews, 54, 1

\bibitem[{Levesque {et~al.}(2005)Levesque, Massey, Olsen, Plez, Josselin,
  Maeder, \& Meynet}]{Levesque2005}
Levesque, E.~M., Massey, P., Olsen, K. a.~G., Plez, B., Josselin, E., Maeder,
  A., \& Meynet, G. 2005, \apj, 628, 973

\bibitem[{{Levesque} {et~al.}(2006){Levesque}, {Massey}, {Olsen}, {Plez},
  {Meynet}, \& {Maeder}}]{Levesque2006}
{Levesque}, E.~M., {Massey}, P., {Olsen}, K.~A.~G., {Plez}, B., {Meynet}, G.,
  \& {Maeder}, A. 2006, \apj, 645, 1102

\bibitem[{{Marco} \& {Bernabeu}(2001)}]{Marco2001}
{Marco}, A., \& {Bernabeu}, G. 2001, \aap, 372, 477

\bibitem[{{Marshall} {et~al.}(2006){Marshall}, {Rajguru}, \&
  {Slosar}}]{Marshall2006}
{Marshall}, P., {Rajguru}, N., \& {Slosar}, A. 2006, \prd, 73, 067302

\bibitem[{{Massey} {et~al.}(2008){Massey}, {Levesque}, {Plez}, \&
  {Olsen}}]{Massey2008}
{Massey}, P., {Levesque}, E.~M., {Plez}, B., \& {Olsen}, K.~A.~G. 2008, in IAU
  Symposium, Vol. 250, IAU Symposium, ed. F.~{Bresolin}, P.~A. {Crowther}, \&
  J.~{Puls}, 97--110

\bibitem[{Massey \& Olsen(2003)}]{Massey2003}
Massey, P., \& Olsen, K. a.~G. 2003, \aj, 126, 2867

\bibitem[{{Massey} {et~al.}(2005){Massey}, {Plez}, {Levesque}, {Olsen},
  {Clayton}, \& {Josselin}}]{Massey2005}
{Massey}, P., {Plez}, B., {Levesque}, E.~M., {Olsen}, K.~A.~G., {Clayton},
  G.~C., \& {Josselin}, E. 2005, \apj, 634, 1286

\bibitem[{{McCall}(2004)}]{McCall2004}
{McCall}, M.~L. 2004, \aj, 128, 2144

\bibitem[{{Meimon} {et~al.}(2005){Meimon}, {Mugnier}, \& {Le
  Besnerais}}]{Meimon2005}
{Meimon}, S., {Mugnier}, L.~M., \& {Le Besnerais}, G. 2005, \josaa, 22, 2348

\bibitem[{{Meimon} {et~al.}(2008){Meimon}, {Mugnier}, \& {Le
  Besnerais}}]{Meimon2008}
---. 2008, \josaa, 26, 108

\bibitem[{{Mendoza}(1967)}]{Mendoza1967}
{Mendoza}, E.~E. 1967, Boletin de los Observatorios Tonantzintla y Tacubaya, 4,
  149

\bibitem[{M\'{e}rand {et~al.}(2005)M\'{e}rand, Bord\'{e}, \& {Coud\'{e} du
  Foresto}}]{Merand2005}
M\'{e}rand, A., Bord\'{e}, P., \& {Coud\'{e} du Foresto}, V. 2005, \aap, 433,
  1155

\bibitem[{{Meynet} \& {Maeder}(2003)}]{Meynet2003}
{Meynet}, G., \& {Maeder}, A. 2003, \aap, 404, 975

\bibitem[{Monnier {et~al.}(2006)Monnier, Pedretti, Thureau, Berger,
  Millan-Gabet, ten Brummelaar, McAlister, Sturmann, Sturmann, \&
  Muirhead}]{Monnier2006}
Monnier, J., {et~al.} 2006, in \procspie, Vol. 6268 (Spie), 62681P

\bibitem[{{Monnier} {et~al.}(2004{\natexlab{a}}){Monnier}, {Berger},
  {Millan-Gabet}, \& {ten Brummelaar}}]{Monnier2004b}
{Monnier}, J.~D., {Berger}, J.-P., {Millan-Gabet}, R., \& {ten Brummelaar},
  T.~A. 2004{\natexlab{a}}, in \procspie, Vol. 5491, 1370

\bibitem[{{Monnier} {et~al.}(2004{\natexlab{b}}){Monnier}, {Millan-Gabet},
  {Tuthill}, {Traub}, {Carleton}, {Coud{\'e} du Foresto}, {Danchi}, {Lacasse},
  {Morel}, {Perrin}, {Porro}, {Schloerb}, \& {Townes}}]{Monnier2004a}
{Monnier}, J.~D., {et~al.} 2004{\natexlab{b}}, \apj, 605, 436

\bibitem[{Monnier {et~al.}(2007)Monnier, Zhao, Pedretti, Thureau, Ireland,
  Muirhead, Berger, Millan-Gabet, {Van Belle}, {Ten Brummelaar}, McAlister,
  Ridgway, Turner, Sturmann, Sturmann, \& Berger}]{Monnier2007}
Monnier, J.~D., {et~al.} 2007, Science, 317, 342

\bibitem[{{Monnier} {et~al.}(2008){Monnier}, {Zhao}, {Pedretti}, {Thureau},
  {Ireland}, {Muirhead}, {Berger}, {Millan-Gabet}, {Van Belle}, {ten
  Brummelaar}, {McAlister}, {Ridgway}, {Turner}, {Sturmann}, {Sturmann},
  {Berger}, {Tannirkulam}, \& {Blum}}]{Monnier2008}
{Monnier}, J.~D., {et~al.} 2008, in \procspie, Vol. 7013

\bibitem[{{Morel} \& {Magnenat}(1978)}]{Morel1978}
{Morel}, M., \& {Magnenat}, P. 1978, \aaps, 34, 477

\bibitem[{{Neugebauer} {et~al.}(1984){Neugebauer}, {Habing}, {van Duinen},
  {Aumann}, {Baud}, {Beichman}, {Beintema}, {Boggess}, {Clegg}, {de Jong},
  {Emerson}, {Gautier}, {Gillett}, {Harris}, {Hauser}, {Houck}, {Jennings},
  {Low}, {Marsden}, {Miley}, {Olnon}, {Pottasch}, {Raimond}, {Rowan-Robinson},
  {Soifer}, {Walker}, {Wesselius}, \& {Young}}]{Neugebauer1984}
{Neugebauer}, G., {et~al.} 1984, \apjl, 278, L1

\bibitem[{{Neugent} {et~al.}(2012){Neugent}, {Massey}, {Skiff}, \&
  {Meynet}}]{Neugent2012}
{Neugent}, K.~F., {Massey}, P., {Skiff}, B., \& {Meynet}, G. 2012, \apj, 749,
  177

\bibitem[{{O'Donnell}(1994)}]{ODonnell1994}
{O'Donnell}, J.~E. 1994, \apj, 422, 158

\bibitem[{{Ohnaka} {et~al.}(2013){Ohnaka}, {Hofmann}, {Schertl}, {Weigelt},
  {Baffa}, {Chelli}, {Petrov}, \& {Robbe-Dubois}}]{Ohnaka2013}
{Ohnaka}, K., {Hofmann}, K.-H., {Schertl}, D., {Weigelt}, G., {Baffa}, C.,
  {Chelli}, A., {Petrov}, R., \& {Robbe-Dubois}, S. 2013, \aap, 555, A24

\bibitem[{{Ohnaka} {et~al.}(2009){Ohnaka}, {Hofmann}, {Benisty}, {Chelli},
  {Driebe}, {Millour}, {Petrov}, {Schertl}, {Stee}, {Vakili}, \&
  {Weigelt}}]{Ohnaka2009}
{Ohnaka}, K., {et~al.} 2009, \aap, 503, 183

\bibitem[{{Ohnaka} {et~al.}(2011){Ohnaka}, {Weigelt}, {Millour}, {Hofmann},
  {Driebe}, {Schertl}, {Chelli}, {Massi}, {Petrov}, \& {Stee}}]{Ohnaka2011}
---. 2011, \aap, 529, A163

\bibitem[{{Perrin}(2003)}]{Perrin2003}
{Perrin}, G. 2003, \aap, 400, 1173

\bibitem[{{Pickles}(1998)}]{Pickles1998}
{Pickles}, A.~J. 1998, \pasp, 110, 863

\bibitem[{{Plez}(2003)}]{Plez2003}
{Plez}, B. 2003, in \pasp, Vol. 298, GAIA Spectroscopy: Science and Technology,
  ed. U.~{Munari}, 189

\bibitem[{{Ragland} {et~al.}(2006){Ragland}, {Traub}, {Berger}, {Danchi},
  {Monnier}, {Willson}, {Carleton}, {Lacasse}, {Millan-Gabet}, {Pedretti},
  {Schloerb}, {Cotton}, {Townes}, {Brewer}, {Haguenauer}, {Kern}, {Labeye},
  {Malbet}, {Malin}, {Pearlman}, {Perraut}, {Souccar}, \&
  {Wallace}}]{Ragland2006}
{Ragland}, S., {et~al.} 2006, \apj, 652, 650

\bibitem[{{Renard} {et~al.}(2011){Renard}, {Thi{\'e}baut}, \&
  {Malbet}}]{Renard2011}
{Renard}, S., {Thi{\'e}baut}, E., \& {Malbet}, F. 2011, \aap, 533, A64

\bibitem[{Rudin {et~al.}(1992)Rudin, Osher, \& Fatemi}]{Rudin1992}
Rudin, L., Osher, S., \& Fatemi, E. 1992, Physica D: Nonlinear Phenomena, 60,
  259

\bibitem[{{Rufener}(1999)}]{Rufener1999}
{Rufener}, F. 1999, VizieR Online Data Catalog, 2169, 0

\bibitem[{{Sargent} {et~al.}(2011){Sargent}, {Srinivasan}, \&
  {Meixner}}]{Sargent2011}
{Sargent}, B.~A., {Srinivasan}, S., \& {Meixner}, M. 2011, \apj, 728, 93

\bibitem[{{Scholz} \& {Takeda}(1987)}]{Scholz1987}
{Scholz}, M., \& {Takeda}, Y. 1987, \aap, 186, 200

\bibitem[{Schwarzschild(1975)}]{Schwarzschild1975}
Schwarzschild, M. 1975, \apj, 195, 137

\bibitem[{Skilling(2006)}]{Skilling2006}
Skilling, J. 2006, Bayesian Analysis, 1, 833

\bibitem[{{Skrutskie} {et~al.}(2006){Skrutskie}, {Cutri}, {Stiening},
  {Weinberg}, {Schneider}, {Carpenter}, {Beichman}, {Capps}, {Chester},
  {Elias}, {Huchra}, {Liebert}, {Lonsdale}, {Monet}, {Price}, {Seitzer},
  {Jarrett}, {Kirkpatrick}, {Gizis}, {Howard}, {Evans}, {Fowler}, {Fullmer},
  {Hurt}, {Light}, {Kopan}, {Marsh}, {McCallon}, {Tam}, {Van Dyk}, \&
  {Wheelock}}]{Skrutskie2006}
{Skrutskie}, M.~F., {et~al.} 2006, \aj, 131, 1163

\bibitem[{Slesnick {et~al.}(2002)Slesnick, Hillenbrand, \&
  Massey}]{Slesnick2002}
Slesnick, C., Hillenbrand, L., \& Massey, P. 2002, \apj, 576, 880

\bibitem[{{Sloan} {et~al.}(2003){Sloan}, {Kraemer}, {Price}, \&
  {Shipman}}]{Sloan2003}
{Sloan}, G.~C., {Kraemer}, K.~E., {Price}, S.~D., \& {Shipman}, R.~F. 2003,
  \apjs, 147, 379

\bibitem[{{Speck} {et~al.}(2000){Speck}, {Barlow}, {Sylvester}, \&
  {Hofmeister}}]{Speck2000}
{Speck}, A.~K., {Barlow}, M.~J., {Sylvester}, R.~J., \& {Hofmeister}, A.~M.
  2000, \aaps, 146, 437

\bibitem[{{Stencel} {et~al.}(1989){Stencel}, {Pesce}, \& {Bauer}}]{Stencel1989}
{Stencel}, R.~E., {Pesce}, J.~E., \& {Bauer}, W.~H. 1989, \aj, 97, 1120

\bibitem[{{Stencel} {et~al.}(1988){Stencel}, {Pesce}, \& {Hagen
  Bauer}}]{Stencel1988}
{Stencel}, R.~E., {Pesce}, J.~E., \& {Hagen Bauer}, W. 1988, \aj, 95, 141

\bibitem[{{Stothers}(2010)}]{Stothers2010}
{Stothers}, R.~B. 2010, \apj, 725, 1170

\bibitem[{{Straizys} {et~al.}(1995){Straizys}, {Kazlauskas}, {Jodinskiene}, \&
  {Bartkevicius}}]{Straizys1995}
{Straizys}, V., {Kazlauskas}, A., {Jodinskiene}, E., \& {Bartkevicius}, A.~.
  1995, VizieR Online Data Catalog, 2157, 0

\bibitem[{Sutton \& Wandelt(2006)}]{Sutton2006}
Sutton, E.~C., \& Wandelt, B.~D. 2006, \apjs, 162, 401

\bibitem[{ten Brummelaar {et~al.}(2005)ten Brummelaar, McAlister, Ridgway,
  Bagnuolo, Turner, Sturmann, Sturmann, Berger, Ogden, Cadman, Hartkopf,
  Hopper, \& Shure}]{tenBrummelaar2005}
ten Brummelaar, T.~a., {et~al.} 2005, \apj, 628, 453

\bibitem[{Thiebaut(2008)}]{Thiebaut2008}
Thiebaut, E. 2008, \procspie, 7013, 70131I

\bibitem[{Thi\'{e}baut \& Giovannelli(2010)}]{Thiebaut2010}
Thi\'{e}baut, E., \& Giovannelli, J. 2010, Signal Processing Magazine, IEEE,
  27, 97

\bibitem[{Tuthill {et~al.}(1997)Tuthill, Haniff, \& Baldwin}]{Tuthill1997}
Tuthill, P., Haniff, C., \& Baldwin, J. 1997, \mnras, 285, 529

\bibitem[{{van Belle} {et~al.}(2009){van Belle}, {Creech-Eakman}, \&
  {Hart}}]{vanBelle2009}
{van Belle}, G.~T., {Creech-Eakman}, M.~J., \& {Hart}, A. 2009, \mnras, 394,
  1925

\bibitem[{{Verhoelst} {et~al.}(2009){Verhoelst}, {van der Zypen}, {Hony},
  {Decin}, {Cami}, \& {Eriksson}}]{Verhoelst2009}
{Verhoelst}, T., {van der Zypen}, N., {Hony}, S., {Decin}, L., {Cami}, J., \&
  {Eriksson}, K. 2009, \aap, 498, 127

\bibitem[{{Wittkowski} {et~al.}(2006){Wittkowski}, {Aufdenberg}, {Driebe},
  {Roccatagliata}, {Szeifert}, \& {Wolff}}]{Wittkowski2006}
{Wittkowski}, M., {Aufdenberg}, J.~P., {Driebe}, T., {Roccatagliata}, V.,
  {Szeifert}, T., \& {Wolff}, B. 2006, \aap, 460, 855

\bibitem[{{Wittkowski} {et~al.}(2004){Wittkowski}, {Aufdenberg}, \&
  {Kervella}}]{Wittkowski2004}
{Wittkowski}, M., {Aufdenberg}, J.~P., \& {Kervella}, P. 2004, \aap, 413, 711

\bibitem[{{Wittkowski} {et~al.}(2012){Wittkowski}, {Hauschildt},
  {Arroyo-Torres}, \& {Marcaide}}]{Wittkowski2012}
{Wittkowski}, M., {Hauschildt}, P.~H., {Arroyo-Torres}, B., \& {Marcaide},
  J.~M. 2012, \aap, 540, L12

\bibitem[{{Woodruff} {et~al.}(2009){Woodruff}, {Ireland}, {Tuthill}, {Monnier},
  {Bedding}, {Danchi}, {Scholz}, {Townes}, \& {Wood}}]{Woodruff2009}
{Woodruff}, H.~C., {et~al.} 2009, \apj, 691, 1328

\bibitem[{{Wright} {et~al.}(2010){Wright}, {Eisenhardt}, {Mainzer}, {Ressler},
  {Cutri}, {Jarrett}, {Kirkpatrick}, {Padgett}, {McMillan}, {Skrutskie},
  {Stanford}, {Cohen}, {Walker}, {Mather}, {Leisawitz}, {Gautier}, {McLean},
  {Benford}, {Lonsdale}, {Blain}, {Mendez}, {Irace}, {Duval}, {Liu}, {Royer},
  {Heinrichsen}, {Howard}, {Shannon}, {Kendall}, {Walsh}, {Larsen}, {Cardon},
  {Schick}, {Schwalm}, {Abid}, {Fabinsky}, {Naes}, \& {Tsai}}]{Wright2010}
{Wright}, E.~L., {et~al.} 2010, \aj, 140, 1868

\bibitem[{{Yang} \& {Jiang}(2012)}]{Yang2012}
{Yang}, M., \& {Jiang}, B.~W. 2012, \apj, 754, 35

\bibitem[{Young {et~al.}(2000)Young, Baldwin, Boysen, Haniff, Lawson, Mackay,
  Pearson, Rogers, St.-Jacques, Warner, Wilson, \& Wilson}]{Young2000}
Young, J.~S., {et~al.} 2000, \mnras, 315, 635

\bibitem[{Zhao {et~al.}(2011)Zhao, Monnier, Che, Pedretti, Thureau, Schaefer,
  ten Brummelaar, M\'{e}rand, Ridgway, McAlister, Turner, Sturmann, Sturmann,
  Goldfinger, \& Farrington}]{Zhao2011}
Zhao, M., {et~al.} 2011, \pasp, 123, 964

\end{thebibliography}
\end{document}